\documentclass[a4paper,10pt,fleqn]{article}

\usepackage{graphicx}
\usepackage{hyperref}
\usepackage{breakurl}
\usepackage{natbib}
\usepackage{setspace}
\usepackage{amsmath}
\usepackage{amssymb}
\usepackage{mathtools}
\RequirePackage[english,iso]{isodate}

\newenvironment{myenumerate}
{\begin{enumerate}
  \setlength{\itemsep}{3.0pt}
  \setlength{\parskip}{0pt}
  \setlength{\parsep}{0pt}}
{\end{enumerate}}

\providecommand{\au}{\ensuremath{{\rm au}}}
\providecommand{\ly}{\ensuremath{{\rm ly}}}
\providecommand{\pc}{\ensuremath{{\rm pc}}}
\providecommand{\kpc}{\ensuremath{{\rm kpc}}}
\providecommand{\yr}{\ensuremath{{\rm yr}}}
\providecommand{\kms}{\ensuremath{{\rm km\,s^{-1}}}}

\providecommand{\as}{\ensuremath{{''}}}
\providecommand{\mas}{\ensuremath{{\rm mas}}}
\providecommand{\muas}{\ensuremath{\mu{\rm as}}}

\providecommand{\vvec}{\ensuremath{{\mathbf v}}}

\providecommand{\xvec}{\ensuremath{{\mathbf x}}}
\providecommand{\rvecdot}{\ensuremath{{\dot{\mathbf r}}}}
\providecommand{\rvec}{\ensuremath{{\mathbf r}}}

\providecommand{\xvecdot}{\ensuremath{{\dot{\mathbf x}}}}
\providecommand{\svec}{\ensuremath{{\mathbf s}}}
\providecommand{\svechat}{\ensuremath{{\hat{\mathbf s}}}}
\providecommand{\svecdot}{\ensuremath{{\dot{\mathbf s}}}}
\providecommand{\svecprime}{\ensuremath{{\mathbf s}^\prime}}
\providecommand{\svechatprime}{\ensuremath{{\hat{\mathbf s}^\prime}}}
\providecommand{\svecdotprime}{\ensuremath{{\dot{\mathbf s}^\prime}}}

\providecommand{\timeobs}{\ensuremath{t_{\rm D}}}
\providecommand{\angdist}{\ensuremath{\rho}}
\providecommand{\rv}{\ensuremath{v}}

\providecommand{\Acirc}{\ensuremath{{\textcircled{\small{A}}}}}
\providecommand{\Bcirc}{\ensuremath{{\textcircled{\small{B}}}}}
\providecommand{\Ccirc}{\ensuremath{{\textcircled{\small{C}}}}}
\providecommand{\Dcirc}{\ensuremath{{\textcircled{\small{D}}}}}

\begin{document}

\vspace*{1cm}

{\LARGE {\bf Lost in space? Relativistic interstellar navigation using an}}

\vspace*{-0.5em}
{\LARGE {\bf astrometric star catalogue}}

\vspace*{0.5cm}
Coryn A.L.\ Bailer-Jones\\
Max Planck Institute for Astronomy\\
K\"onigstuhl, 69117 Heidelberg, Germany\\
\url{http://www.mpia.de/homes/calj}\\

Accepted to {\em Publications of the Astronomical Society of the Pacific} on 2 June 2021.\\
Originally submitted for publication 13 October 2020.\\
Minor revisions 18 March 2021 (arXiv v1) and 12 May 2021 (arXiv v2).

%%%%%%%%%%%%%%%%%%%%%%%%%%%%%%%%%%%%%%%%%%%%%%%%%%%%%%%%%%%%%%%%%%%%%%%
\section*{Abstract}

The exploration of interstellar space will require autonomous navigation systems that do not rely on tracking from the Earth.  Here I develop a method to determine the 3D position and 3D velocity of a spacecraft in deep space using a star catalogue.  As a spacecraft moves away from the Sun, the observed positions and velocities of the stars will change relative to those in a Earth-based catalogue due to parallax and aberration.  By measuring just the angular distances between pairs of stars, and comparing these to the catalogue, we can infer the coordinates of the spacecraft via an iterative forward-modelling process.  I perform simulations with existing star catalogues to demonstrate the method and to compute its performance.  Using the 20 nearest stars and a modest angular distance measurement accuracy of 1\as, the position and velocity of a spacecraft light years from the Sun moving at relativistic speeds can be determined to within 3\,\au\ and 2\,\kms\ respectively.  These accuracies improve linearly with the measurement accuracy, e.g.\ with angles measured to 1\,\mas\ the navigation accuracy is 1000 times better.  Performance can also be improved using more stars, or by including onboard measurements of the stars' radial velocities, as these too are affected by the spacecraft's position and motion.
%The method also estimates the uncertainties in its inferred coordinates, and these are found to be reasonably reliable.
%Although pulsar timing measurements potentially offer a more accurate means to navigate, stellar positional or radial velocity navigation can serve as an independent check.

\vspace*{1em}
Keywords: navigation -- interstellar travel -- astrometry -- relativity -- parallax -- aberration

%%%%%%%%%%%%%%%%%%%%%%%%%%%%%%%%%%%%%%%%%%%%%%%%%%%%%%%%%%%%%%%%%%%%%%%
\section{Introduction}

How do we determine our position and velocity in space?  For spacecraft near to the Earth this is routinely done with radar tracking using ground-based antenna.  A signal sent to and then returned by a spacecraft can be used to determine the spacecraft's distance via the time delay and its radial velocity via the Doppler effect.
Accuracies of 1\,m and 1\,mm\,s$^{-1}$ respectively are routinely achieved in this way.
%%% N01
The direction to the spacecraft (two-dimensional position) is also determined from the antenna pointing, albeit less accurately. By tracking the spacecraft in this way over time, its orbit may be computed. This can be done more quickly or more accurately by tracking simultaneously with two or more ground stations to enable a triangulation.

Radar tracking is less accurate for deep space, and so is augmented by the Delta-Differential One-Way Ranging ($\Delta$DOR) technique (e.g.\ James et al.\ 2009).  This uses two widely-spaced ground stations to receive a signal sent by the spacecraft. The time-delay between the receipt of the signals establishes the angular position of the spacecraft, the accuracy of which is then improved by repeating this measurement for a quasar that lies within a few degrees of the line-of-sight. This provides the direction to the spacecraft relative to the quasar, the position of which is already known through earlier observations. An accuracy of the order 10\,nrad (2\,mas) may be achieved (Iess et al.\ 2014), which corresponds to a transverse positional accuracy of 1.5\,km for a spacecraft 1\,\au\ ($1.5\times10^{8}$\,km) from the Earth.

Beyond the solar system, spacecraft will be too distant to rely on Earth-based tracking.  When travelling to the nearest stars, signals will be far too weak and light travel times will be of order years. An interstellar spacecraft will therefore have to navigate autonomously, and use this information to decide when to make course corrections or to switch on instruments. Such a spacecraft needs to be able to determine its position and velocity using only onboard measurements.  In principle this can be done by integrating over internal measurements from clocks, gyroscopes, and accelerometers (e.g.\ Hoag \& Wrigley 1975), but in practice this would be neither accurate nor reliable enough for mission durations of decades.
%Moreover, as positions and velocities are only meaningful relative to external bodies, we should anyway make measurements relative to some external reference frame.

Pulsars are a well-studied proposed solution to the problem of deep space navigation (e.g.\ Shemar et al.\ 2016, Becker et al.\ 2013). Pulsars are rapidly rotating neutron stars that emit narrow radio and X-ray pulses at very stable and well-defined rates, with periods ranging from milliseconds to seconds.  By measuring the arrival time of a pulse from a single pulsar and comparing it to the expected arrival time, we can determine the spacecraft position relative to its expected position along the line-of-sight to the pulsar. As pulses are indistinguishable, this only establishes the position as somewhere on one of an infinite number of planes that are separated by a distance $cT$, where $c$ is the speed of light and $T$ is the period of the pulsar. By repeating this measurement for multiple pulsars in different directions, we can break the degeneracies and determine the three-dimensional position of the spacecraft.
This method is analogous (but not identical) to global positioning systems on Earth, which use satellites as opposed to pulsars as navigation beacons.
The accuracy of pulsar navigation is set, among other things, by the accuracy with which the pulsars' periods and directions have been determined in advance (on the Earth).  Simulations by Shemar et al.\ (2016) predict that a spacecraft position could be determined to an accuracy of 2\,km\ for a spacecraft up to 30\,au from the Earth.
Errors in the timing model due to dispersion by the interstellar medium may reduce this accuracy at light year distances, however.
Velocities could be estimated from Doppler shifts of the pulsar signals, either from their wavelength or their frequency of arrival. One could also use two position determinations separated by a short period of spacecraft time to give the proper velocity.
% this gives the proper velocity w, related to the coordinate velocity beta by 1/beta^2= 1 + 1/w^2 (in units of c).
%%% N02

Another way to determine spacecraft position in deep space is via direct triangulation of stars (e.g.\ Moskowitz \& Devereu 1968, Hoag \& Wrigley 1975). If the 3D positions of a set of stars relative to some reference frame are known, then the observed parallactic shift of these positions can be used to compute the position of the spacecraft in the reference frame. In reality stars are moving relative to the spacecraft, and due to the finite speed of light this introduces an apparent shift of the stars' positions due to aberration. The size of this aberrational shift depends on the relative velocity, and so we could exploit this to determine the velocity of the spacecraft (e.g.\ Butkevich \& Klioner 2008). Indeed, if we use sources such as quasars that are effectively infinitely far away, and so exhibit negligible parallactic shifts, or if we limit the navigation to spacecraft near the solar system, then we could use the aberration to infer just the velocity (Calabro' 2011, Christian 2019).

In this paper I develop and test a scheme for determining the 3D position and 3D velocity (6D coordinates) of a spacecraft from measurements of the angular positions and/or radial (Doppler) velocities of stars.  These are the only plausible measurements of stellar positions and velocities that can be made both quasi-instantaneously and without reference to an external system.  As it would be difficult in practice to establish the absolute 2D angular coordinates of stars onboard a spacecraft, positional measurements are limited to one-dimensional angular distances between pairs of stars, as can be made with a sextant, for example.

%As the mathematical relationship between the spacecraft coordinates and the measurements is nonlinear, there is no direct solution.  I adopt a forward modelling procedure to predict the measurements from assumed spacecraft coordinates, and then use a Monte Carlo method to iterate.

The navigation scheme uses a catalogue of 6D coordinates of the stars in some reference system.  Astrometric catalogues provide five of these six coordinates -- latitude, longitude, distance, and two transverse velocities -- with the sixth provided by radial velocity measurements from optical spectrographs. Space astrometry and spectroscopy with the Gaia spacecraft now provide such catalogues at sub-milliarcsecond and \kms\ accuracy for millions of stars (Gaia Collaboration 2018). The principle of constructing these catalogues is in some sense the inverse of navigation: the positions and velocities of stars are determined in the reference system -- usually the International Celestial Reference System (ICRS) --
using the known position and velocity of the observer relative to this reference system.  The scheme I present here takes into account the parallax, space motion, and aberration of the stars.

To reach the nearest stars within a human lifetime, a spacecraft will need to travel at relativistic speeds, and so the calculations are done here using special relativity. General relativistic light bending can be neglected: once we are more than 100\,au\ from a solar mass star the effect is below 0.1\,\mas, and so below the best measurement accuracies we will consider.
% 4*GM/(c^2 r ) = (3600/conv)* (4*GG*msun/cc^2)/(100*au) as
Its accommodation would not, however, change the principle of the scheme presented.

I describe the scheme in section~\ref{sec:method}. Using simulations of the spacecraft position and velocity, together with simulated, noisy measurements of a real star catalogue (section~\ref{sec:setup}), I investigate the performance as a function of various parameters in section~\ref{sec:results}. These results are further discussed in section~\ref{sec:conclusions} along with some practical considerations for implementing this approach.
Although the concept of interstellar navigation with a star catalogue is not new, this is the first time, to the best of my knowledge, that it has been presented in detail and tested via simulations. 

%%%%%%%%%%%%%%%%%%%%%%%%%%%%%%%%%%%%%%%%%%%%%%%%%%%%%%%%%%%%%%%%%%%%%%%
\section{Navigation method}\label{sec:method}

I first outline in section~\ref{sec:principle} the general principle of the navigation procedure before providing more technical details in section~\ref{sec:technical}. 

\subsection{Principle}\label{sec:principle}

Our starting point is a catalogue of 3D positions and 3D velocities of bright stars within a few tens of parsec from the Sun.  These coordinates are in the ICRS at some reference time. We may neglect the uncertainties in these coordinates, because as we will see below, the uncertainties of the measurements we can expect to make onboard a spacecraft are usually much larger than the uncertainties in modern star catalogues. The origin of the ICRS is the solar system barycentre (SSB).

Imagine for a moment that we had a 2D angular coordinate grid -- the latitude (declination, Dec) and longitude (right ascension, RA) -- of the ICRS imprinted on the sky.  As our spacecraft moves, any star will change its position on this grid due to parallax.
A star 1\,\pc\ away will be displaced by 1\,\as\
% (arcsecond $=\pi/(180\times3600$\,rad)
for every astronomical unit
% ($\au = 1.5\times10^{11}$\,m)
that the spacecraft moves, by definition of the parsec ($\pc = 180\times3600/\pi\,\au \simeq 206\,000\,\au$). 
As our spacecraft will move to much larger distances -- up to several \pc\ -- the displacement will be much larger and easily measurable.

Of course, we do not have a fixed grid imprinted on the sky, and so we cannot directly measure the RA and Dec of the stars.\footnote{We can consider quasars to be more or less fixed on the sky, and so to provide a realization of the ICRS that remains fixed at the microarcsecond level. But quasars are faint, plus making independent measurements of RA and Dec would be difficult in practice, so we do not rely on this.}
We instead measure the angular distances between pairs of stars. These angular distances will also change due to parallax, yet they can be measured without reference to an external coordinate system. This is the principle of a sextant, and is essentially how the Gaia satellite works (Gaia Collaboration 2016).
Commercial star trackers on satellites can measure these angular distances to an accuracy of around 1\as.
When only considering parallax, three angular distances in principle suffice to determine the 3D position of our spacecraft.

Parallax is not the only effect we must consider, however. Because our spacecraft is moving relative to the stars, their apparent positions will change also because of aberration, a consequence of the finite speed of light. Aberration is a large effect: for an observer moving perpendicular to the line-of-sight to a star at a velocity $u$,
% relative to that star,
the position of the star will appear to shift by $\arcsin(u/c)$ radians compared to its position observed when at rest.  This is readily observable from the Earth: the apparent position of a star changes by 41\as\ over the course of a year due to the orbit of the Earth around the Sun, because this changes the velocity of the Earth relative to any star by $\pm 30\,\kms$.  In our application, as we know the velocity of the star relative to the SSB, the aberration encodes information about the velocity of the spacecraft relative to the SSB.  A single measurement of aberration cannot solve for the 3D velocity, not least because the observed position of the star is also shifted by the parallax.  But if we measure the positions of many stars -- by which we mean the angular distances between many stars -- then we can expect to untangle the effects of parallax and aberration to determine the 3D position and 3D velocity of our spacecraft.
Given the nonlinearity of the expressions involved, we solve this via a forward-modelling approach, in which we compare the measured angular distances with those expected based on an assumed position and velocity, then iteratively update this to minimize the differences between the measured and expected angular distances.

Stars move relative to the SSB, so we have to compute their expected positions (in the ICRS) using their known velocities and
the time of observation. Due to the finite speed of light, the position of a star when observed is not the same as its actual position at that time.
Thus at each step in our iterative procedure we will use the computed distance between the spacecraft and a star to correct the star's observed position for the light travel time.\footnote{We will see in section~\ref{sec:technical} that only differences in light travel times -- and hence changes in the stars' distances -- are relevant, so this still works when stars are arbitrarily far away.}
Since our spacecraft is moving relativistically in general -- and we won't assume to know how its speed has varied since it left the SSB -- we cannot assume to know with any accuracy the time in the SSB frame at the moment of observation. I therefore adopt this time as the seventh parameter
% -- in addition to 3D position and 3D velocity --
in the navigation problem and solve for this as part of the iterative process.
%\footnote{For simplicity I assume that all measurements are taken at the same time. If not we would still only need one time parameter, because we could measure the spacecraft time between our measurements.}
%%% N03

The set of measurements we have so far considered is the angular distances (great circle angles) between $N-1$ stars and an arbitrarily chosen reference star (e.g.\ the Sun).  As we need to solve for seven parameters we expect to need of order ten stars, although we will analyse a range of numbers in section~\ref{sec:results}. In principle we could obtain more measurements by measuring all $N(N-1)/2$ possible angular distances between these stars, but this would introduce some correlations in the data.
The onboard measurements might also be easier if one field-of-view of the sextant is kept pointed at the same reference star.

The other set of measurements we can consider is radial velocities of the stars, which we can measure via the Doppler shift using a spectrograph. A calibration lamp on the spacecraft can determine the radial velocities relative to the spacecraft without need for an external reference. We will investigate in section~\ref{sec:results} the quality of using angular distances and/or radial velocities of various degrees of accuracy.

We will not consider including measurements of the distances or transverse velocities of the stars, as it is unfeasible to measure these from the spacecraft. Parallaxes and proper motions are usually inferred from the change in angular position measurements when moving over a known baseline, yet that is precisely what we do not know.

\subsection{Technical details}\label{sec:technical}

\begin{table}
\caption{Summary of the steps in the navigation scheme. Symbols are defined in Table~\ref{tab:definitions}.
\label{tab:inference_procedure}
}
\vspace*{0.5em}
\par\noindent\rule[0.5em]{\textwidth}{0.25pt}
The ICRS positions $\{\rvec\}$ and velocities $\{\rvecdot\}$ of the stars at time zero are known.
\begin{myenumerate}
\item Initialize with estimates of the spacecraft position \xvec, velocity \xvecdot, and measurement time \timeobs.
\item Predict the expected unit position vector \svechat, and velocity \svecdot, of each star.
  \begin{myenumerate}
  \item Compute \svecprime, the expected position vector of the star relative to the fictitious observer,
    from \rvec, \rvecdot, \xvec, and \timeobs\ (eqn.~\ref{eqn:svecprime}).
  \item \svecdotprime = \rvecdot, as the fictitious observer is stationary relative to the SSB (ICRS).
  \item Compute \svechat\ from \svechatprime\ and \xvecdot\ using the expression for aberration (eqn.~\ref{eqn:aberration}).
  \item Compute \svecdot\ from \svecdotprime\ and \xvecdot\ using the Lorentz velocity transformation (eqn.~\ref{eqn:lorentz}).
  \end{myenumerate}
\item Compute the expected data for each star.
  \begin{myenumerate}
  \item Compute the angular distances between each star and the reference star from $\{\svechat\}$.
  \item Compute the radial velocities for each star from $\{\svecdot\}$.
  \end{myenumerate}
\item Evaluate the likelihood from the measured data, the expected data, and the expected uncertainties.
\item Use the likelihood to update the spacecraft parameters (with MCMC), and iterate from step 2.
\end{myenumerate}
\par\noindent\rule[1em]{\textwidth}{0.25pt}
\end{table}

I now describe in detail the steps in the inference procedure, summarized also in Table~\ref{tab:inference_procedure}.  The goal is to infer the seven spacecraft parameters, namely the 3D position, 3D velocity, and time of the measurement, relative to the SSB. As we adopt a forward modelling approach, much of this involves computing the expected positions and velocities of the stars from the adopted spacecraft parameters.

\begin{figure}
\begin{center}
\includegraphics[width=0.5\textwidth, angle=0]{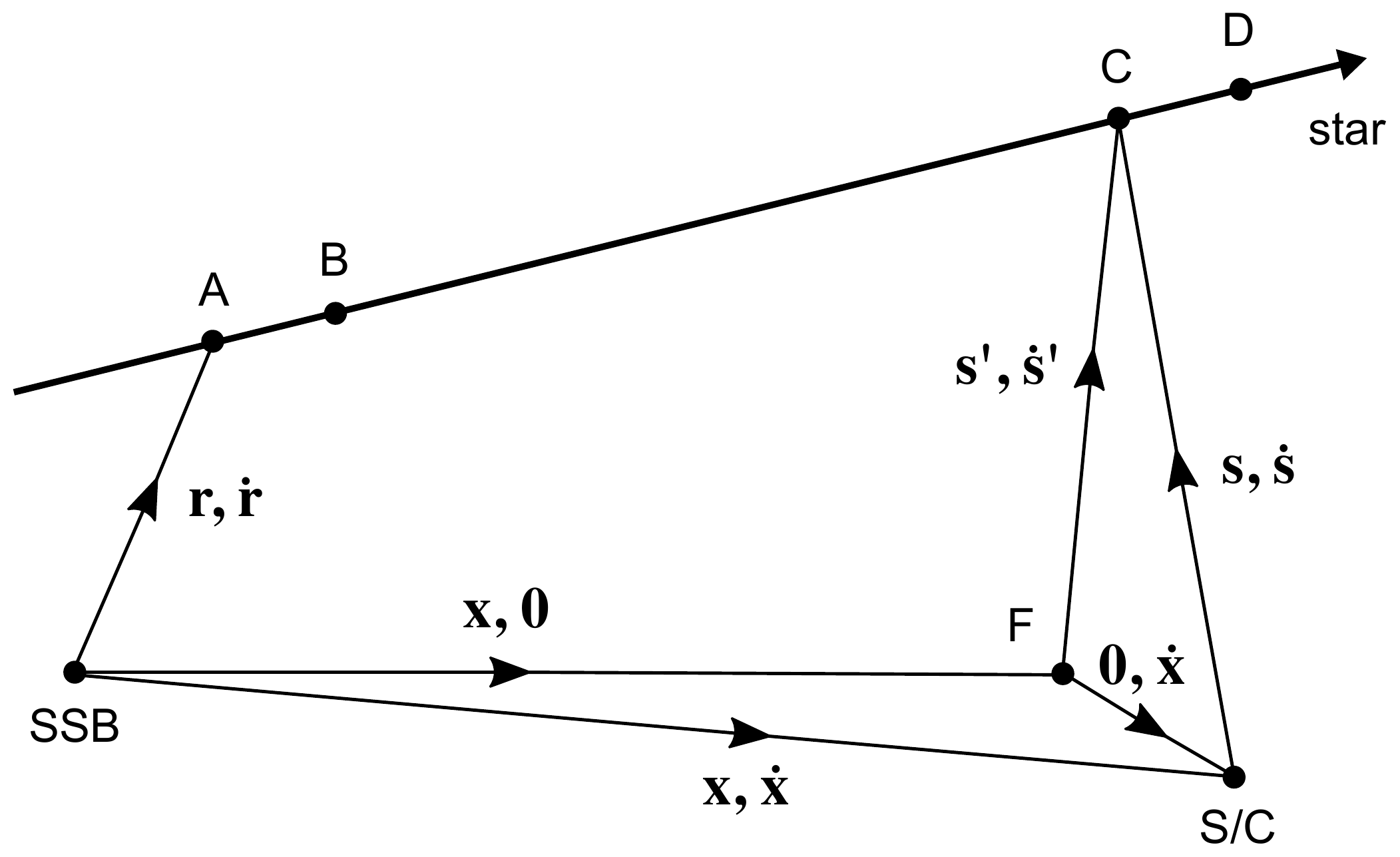}
\caption{
  Schematic representation of the navigation problem as seen from the
  solar system barycentre (SSB), showing the path of one star (line at the top), and the positions
  of the SSB, the spacecraft (S/C), and the fictitious observer (F), defined as one who is at the instantaneous position of the spacecraft but is at rest relative to the SSB.
  %\Acirc, \Bcirc, \Ccirc\ and \Dcirc\
  A, B, C, and D are positions of the star at various times.
  The two symbols next to the vectors (lines with arrows in the centres), e.g.\ \rvec, \rvecdot, denote the position and velocity respectively of the point at the end of the vector relative to the point at the beginning.
See Table~\ref{tab:definitions} for a list of definitions.
  S/C is drawn offset from F for clarity; in reality they are at the same position (but different velocities).
\label{fig:situation}
}
\end{center}
\end{figure}

\begin{table}
\begin{center}
  \caption{Definitions of symbols. All positions and velocities are at SSB time \timeobs,
%%% N04
except for \rvec\ and \rvecdot, for which a subscript A, B, C, or D is used in the text to denote the
SSB time when the star is at these positions indicated in Figure~\ref{fig:situation}.
\label{tab:definitions}
}
\vspace*{1em}
\begin{tabular}{ll}
\hline
symbol & meaning  \\
\hline
  \xvec  & 3D position of the spacecraft relative to the SSB \\
  \xvecdot  & 3D velocity of the spacecraft relative to the SSB \\
  \timeobs &  time in the SSB of the measurement \\
  \rvec  &  3D position of a star relative to the SSB \\
  \rvecdot  &  3D velocity of a star relative to the SSB \\
  \svec  &  3D position of a star relative to the spacecraft\\
  \svechat  & unit vector corresponding to \svec \\
  \svecdot  & 3D velocity of a star relative to the spacecraft\\
  \svecprime  & 3D position of a star relative to the fictitious observer\\
  \svechatprime  & unit vector corresponding to \svecprime \\
  \svecdotprime & 3D velocity of a star relative to the fictitious observer\\
  \angdist & angular distance between two stars as seen from the spacecraft \\
  \rv & radial velocity of a star relative to the spacecraft \\
   $c$ & speed of light\\
\hline
\end{tabular}
\end{center}
\end{table}

The position and velocity relations between the SSB, the spacecraft, and an example star are shown schematically in Figure~\ref{fig:situation}. Table~\ref{tab:definitions} defines the symbols.  To ease the transformation of positions and velocities, I introduce the concept of the {\em fictitious observer} following the approach of Klioner (2003). The fictitious observer, denoted F in Figure~\ref{fig:situation}, is defined as one who is at the same instantaneous position as the spacecraft (S/C in the figure), but has zero velocity relative to the SSB.
%%% SSB and F can have x,y,z axes aligned.

\subsubsection{Star motions and light travel time: \svecprime\ and \svecdotprime\ from \rvec, \rvecdot, \xvec, and \timeobs}\label{sec:star_motions}

The first step is to determine the position and velocity of a star as seen by the fictitious observer.

Our star catalogue lists the 3D positions and 3D velocities of $N$ stars relative to the SSB at some SSB reference time which we define as $t=0$.  I assume that the stars move on unaccelerated paths relative to the SSB, so this catalogue completely describes their future motion. The stellar velocities are non-relativistic (typically tens of \kms) so their motion is treated classically. See Butkevitch \& Lindegren (2014) for a more detailed analysis.
%%% N05

A star's recorded position in the catalogue is point \Acirc\ in Figure~\ref{fig:situation}. I denote this as $\rvec_{\rm A}$.  The time (in the SSB) required for the photons to reach the SSB is $|\rvec_{\rm A}|/c$, at which point the star has moved to \Bcirc. This position relative to the SSB is \begin{equation} \rvec_{\rm B} = \rvec_{\rm A} + \rvecdot\,\frac{|\rvec_{\rm A}|}{c} \ .  \end{equation} The photons that the spacecraft receives to make its measurements are emitted by the star when it is at \Ccirc, but the star will be at \Dcirc\ by the time these photons arrive at the spacecraft.  This time of arrival, \timeobs, is the same time it takes the star to move from \Bcirc\ (where it was at $t=0$) to \Dcirc. Therefore \begin{equation}
  \rvec_{\rm D} = \rvec_{\rm B} + \rvecdot\,\timeobs \ .
\end{equation}
The apparent position of the star relative to the fictitious observer at the moment of measurement is \svecprime.
The time taken for the light to travel from the star to the fictitious observer is therefore $|\svecprime|/c$.
By construction this equals the time taken for the star to move from \Ccirc\ to \Dcirc, so
\begin{equation}
  \rvec_{\rm C} = \rvec_{\rm D} - \rvecdot\,\frac{|\svecprime|}{c} \ .
\end{equation}
It follows from the definitions of the position vectors (see Figure~\ref{fig:situation}) that the position of the star as seen by the fictitious observer is
\begin{equation}
  \svecprime = \rvec_{\rm C} - \xvec \ . 
\end{equation}
Putting together the previous four equations we get
\begin{equation}
\svecprime = \rvec_{\rm A} - \xvec + \underbrace{\rvecdot\left( \frac{|\rvec_{\rm A}|}{c} - \frac{|\svecprime|}{c} + \timeobs \right)}_{\rm Q} \ .
\label{eqn:svecprime}
\end{equation}
$\rvec_{\rm A}$ and \rvecdot\ come from our catalogue, and 
at any step in our inference process we are in possession of assumed values of \xvec\ and \timeobs. We can therefore solve for \svecprime. This is complicated by the fact that the magnitude of \svecprime\ appears separately in this equation, but we can easily overcome this by solving iteratively (a direct solution is also possible).
%%% similar direct solution is Eqn. 33 of Butkevitch & Lindegren (2014).
As the stars are not very distant and move at a small fraction of the speed of light, the magnitude of the term marked Q in equation~\ref{eqn:svecprime} is small compared to the other terms: it is typically of order $50\,\kms \times 5\,\pc / c \sim 0.001\,\pc$, compared to a few \pc\ for the other terms. I therefore initialize the iterations by setting Q\,$=0$. The magnitude of \svecprime\ converges to better than 1 metre within five iterations for a wide range of plausible starting values.

%%% N06
As we are dealing with a relativistic spacecraft we need to be clear about the treatment of time.  The spacecraft and fictitious observer are at the same point in space so we can consider the measurement -- the reception of photons from the star -- to be simultaneous in these two frames.
%%% This can in principle be recorded in SSB time because a clock at the fictitious observer is at rest with respect to the SSB.
The actual time that a clock on the spacecraft records at this moment is unimportant.
%%% The actual times they each record on their own clocks depend on when their clocks were synchronized and how they have moved relative to each other since. They will not in general be equal, nor will they in general be equal to the SSB time. But this is unimportant because for the navigation problem we are not interested in the time onboard the spacecraft (which can anyway be measured by a clock on the spacecraft).
All we need to know (or rather infer) is the time in the SSB frame at the moment the star's photons reach the spacecraft (\timeobs), because this is what is required to accommodate for the motion of the star.\footnote{We are allowed to talk about the SSB time of this event, because the fictitious observer, who is at rest with respect to the SSB, can in principle record SSB time.}
In this sense \timeobs\ is an auxiliary parameter. It only enters into the inference via equation~\ref{eqn:svecprime}, and here only as a product with the star's space velocity. This is also the case for the light travel times, which are relevant only because the star is moving.  If stars didn't move relative to the SSB we wouldn't have to include \timeobs\ in the inference.  As noted above, Q in equation~\ref{eqn:svecprime} is usually small, meaning the stars' motions have little impact on the inference in most cases.
% (and correspondingly cannot be inferred with high accuracy).

It is worth noting further that light travel time only enters into equation~\ref{eqn:svecprime} as the {\em difference} in the light travel time between the star and the SSB, and the star and the spacecraft. The total time for light to travel from the star does not play a role.  In particular, for stars that are effectively infinitely far away (so have no parallax), we will have $\svecprime = \rvec_{\rm A}$; light travel time effects disappear and \timeobs\ is no longer relevant.
%%% N07
In these cases, measurements of the star's positions of course no longer tell us anything about the spacecraft position.

We have now computed the positions of the stars relative to the fictitious observer.  Computing their velocities relative to the fictitious observer is trivial, because the observe is stationary relative to the SSB. Hence \svecdotprime = \rvecdot.

\subsubsection{Aberration: \svechat\ from \svechatprime\ and \xvecdot}

We now know \svecprime, and therefore its corresponding unit vector \svechatprime, the direction to the star relative to the fictitious observer. To transform from the fictitious observer to the spacecraft frame we need to correct for aberration.
With
\begin{equation}
  \gamma \,=\, \left(1 - \frac{|\xvecdot|^2}{c^2} \right)^{-1/2} \ ,
  \label{eqn:gamma}
\end{equation}
the direction to the star in the spacecraft frame is given by
\begin{equation}
  \svechat \,=\, \left(\svechatprime + \left[\frac{\gamma}{c} +
      (\gamma-1) \frac{\xvecdot \cdot \svechatprime}{|\xvecdot|^2} \right] \xvecdot \right) 
  \left( \gamma + \left[ 1 + \frac{\xvecdot \cdot \svechatprime}{c} \right] \right )^{-1} \ .
  \label{eqn:aberration}
\end{equation}
This is equation 10 of Klioner (2003), where in his notation I set $\vvec = \xvecdot$ because I am neglecting distortions from space-time curvature. Aberration is independent of distance, and so it transforms the unit vector \svechatprime\ to the unit vector \svechat\ (rather than \svecprime\ to \svec). As we are not attempting to measure the distance to any star, this is sufficient.

\subsubsection{Lorentz velocity transformation: \svecdot\ from \svecdotprime\ and \xvecdot}

To transform the velocity of the star from the fictitious observer to the spacecraft frame we use the Lorentz transformation, which in vector form is
\begin{equation}
  \svecdot \,=\, \left( \frac{\svecdotprime}{\gamma} - \xvecdot \left[ 1 -  \frac{\xvecdot \cdot \svecdotprime}{c^2}
      \left(\frac{\gamma}{\gamma+1}\right) \right] \right)
  \left( 1 - \frac{\xvecdot \cdot \svecdotprime}{c^2} \right)^{-1}
\label{eqn:lorentz}
\end{equation}
% text book equation
with $\gamma$ defined by equation~\ref{eqn:gamma}.

\subsubsection{Expected data}

We have now evaluated the expected direction to the star \svechat, and the expected velocity of the star \svecdot, relative to the spacecraft.
% (in Cartesian coordinates).
This is done for all $N$ stars.
We then
% convert to spherical polar coordinates and
compute the angular distances between each star and the reference star using the haversine formula. If $\theta$ is the latitude (Dec) and $\phi$ the longitude (RA) of a star, and $(\theta_r,\phi_r)$ the coordinates of the reference star, then the angle between them is
\begin{equation}
  \angdist \,=\, 2\arcsin\left( \left[ \sin^2\!\left(\frac{\theta-\theta_r}{2}\right) +
      \cos(\theta)\cos(\theta_r)\sin^2\!\left(\frac{\phi-\phi_r}{2}\right)  \right]^{1/2} \right) \ .
\label{eqn:haversine}
\end{equation}
For $N$ stars we measure $N-1$ angular distances.

The other measurement we will consider is the radial velocity \rv. The expected value of this is the radial component of \svecdot.
% once this is converted to spherical polar coordinates.
We have $N$ of these, but in the interests of keeping a consistent set of measurements we will not use the radial velocity of the reference star.

It is important to realise that we only make use of the sky coordinate systems in the spacecraft and fictitious observer frames
for the purpose of computing the expected data from the star catalogue data and spacecraft parameters. We do not rely on knowing these coordinate systems when making either the angular distance or radial velocity measurements, as these are entirely local to the spacecraft.
%The SSB frame is the basis for the star catalogue, and is the frame relative to which we determine the spacecraft position and velocity.

\subsubsection{Measured data}

Lacking an actual spacecraft to provide measurements, I simulate the measured data by adding Gaussian random noise to the expected data.  Denoting the measured and expected values of the data with the subscripts ``meas'' and ``exp'', and denoting the standard deviation of the noise in the angular distance as $\sigma_\angdist$ and in the radial velocity as $\sigma_\rv$, the simulated measurements are
\begin{alignat}{2}
  \angdist_{\rm meas} \,&=&\, \angdist_{\rm exp} + {\cal N}(0,\sigma_\angdist) \label{eqn:noise_angdist} \\
  \rv_{\rm meas}           \,&=&\, \rv_{\rm exp} + {\cal N}(0,\sigma_\rv) \label{eqn:noise_rv}
\end{alignat}
where  ${\cal N}(0,\sigma)$ indicates a Gaussian random variable of zero mean and standard deviation $\sigma$.

To compare these measurements to their expected values, we assume that we
can crossmatch the observed stars to those in the catalogue
from the spacecraft.  If the parallax and/or aberration are large, this is not as simple as it seems, even though the stars are bright.  But as long as we have regularly tracked the stars since departing the Earth, this is a solvable problem.

\subsubsection{Likelihood}

To quantify the deviation of the measured data from the expected data I use the likelihood.  This is formally the probability density of the measured data given the model parameters, and is specified by the measurement noise model.  Assuming the measurements to be independent, the likelihood for each can be approximated as a one-dimensional Gaussian distribution with standard deviation equal to the expected standard deviation of the noise.  For simplicity we will take this standard deviation as the same for all stars. This would be appropriate if, as it likely the case, it is set by systematic rather than random errors.  The log likelihood for the $N-1$ angular distances can then be written as
\begin{equation}
  \ln L_\angdist \,=\,
  - \sum_{i=1}^{N-1}\left( \frac{1}{2}\left[\frac{\angdist^{(i)}_{\rm meas}-\angdist^{(i)}_{\rm exp}}{\sigma_\angdist} \right]^2 + \ln \sigma_\angdist \right)
\label{eqn:likelihood_angdist}
\end{equation}
and for the radial velocities as
\begin{equation}
  \ln L_\rv \,=\,  
  - \sum_{i=1}^{N-1}\left( \frac{1}{2}\left[\frac{\rv^{(i)}_{\rm meas}-\rv^{(i)}_{\rm exp}}{\sigma_\rv} \right]^2 + \ln \sigma_\rv \right) 
\label{eqn:likelihood_rv}
\end{equation}
both to within an additive constant. When we use both types of measurement in the inference the log likelihood is the sum of these two likelihoods. In principle we could add to this the logarithm of a prior probability density over the model parameters to get the log posterior. But we will see that in most cases the data are quite informative, so I do not include an explicit prior. Implicitly the prior is uniform in the parmeters.

In the simulations in section~\ref{sec:results} we will adopt a nominal angular distance measurement accuracy (``error'') of order 1\as. This is several orders of magnitude worse than what Gaia is now achieving for position and parallax (and proper motions are as precise as $10^{-5}\,\mas/\yr$; Gaia Collaboration 2018).
Ground-based radial velocity uncertainties can be as low as a few m\,s$^{-1}$.
We are therefore justified in neglecting the catalogue uncertainties in the likelihood. If we needed to introduce these, we could simply inflate $\sigma_\angdist$ and $\sigma_\rv$ accordingly.

\subsubsection{Markov Chain Monte Carlo (MCMC)}

I use MCMC to iteratively explore the 7-dimensional parameter space to infer the model parameters and uncertainties therein.  MCMC updates the model parameters based on changes in the likelihood.  Here I use my R implementation of the method of Goodman \& Weare (2010) and Foreman-Mackey et al.\ (2013), which moves a set of ``walkers'' in parallel through parameter space. The size of the update to each walker depends on the distribution of all the other walkers. This method has the advantage over the more traditional Metropolis method that we do not need to fix a typical step size, so it is particularly appropriate when we have little idea of the parameter differences that can be resolved by the likelihood.

Any MCMC must be initialized at some values of the parameters. It seems reasonable to suppose that during the course of its interstellar journey the spacecraft would know its position, velocity, and the SSB time to within 10\% of their true values, by virtue of having regularly determine these parameters (with the method described here) since it departed from the Earth.\footnote{10\% is conservatively large in practice; a more precise initialization would permit faster convergence.}
I therefore initialize the seven parameters of each walker by drawing from a distribution that is uniform over the range 0.9 to 1.1 times the true value.
%As the position and velocity parameters are computed in Cartesian coordinates, and so can be negative, their true values could be arbitrarily close to zero. I therefore impose the additional restriction that the initialization range is never narrower than $\pm100\,\au$ for the three position parameters and $\pm100\,\kms$ for the three velocity parameters.  In practice this additional restriction is only triggered for the non-relativistic simulations described in section~\ref{sec:setup}.

After some testing I settled on using 40 walkers. 500 iterations of burn-in (for each walker) are generally sufficient for the walkers to migrate from the initial position to the region of non-tiny likelihoods, although for the simulations with the highest accuracy measurements -- and so the narrowest likelihood functions -- I increase this, up to 2500 iterations. From there I allow the walkers to sample for a further 800 iterations each.
After discarding the burn-in I thin the remaining samples by retaining every 5th iteration from every 5th walker, to leave a final set of 1280 likelihood samples. The time series of these is referred to as a ``chain''. To estimate the spacecraft parameters, I take the median of the samples, and to estimate their uncertainties (not errors!) I take the standard deviation of the samples.

\section{Set-up for the simulations}\label{sec:setup}

\subsection{Star catalogue}

I initially tested the navigation scheme using simulated star catalogues, but the results in section~\ref{sec:results} use the actual solar neighbourhood.  For this I use the Hipparcos astrometric catalogue (Perryman 1997) supplemented by radial velocities from the Simbad astronomical database (Wenger et al.\ 2000).  Together these provide the six position and velocity coordinates in spherical polar coordinates, which I convert to Cartesian for the vector calculations.  Hipparcos actually provides a parallax, not a distance, and because this is noisy its simple reciprocal is not an optimal estimate of distance (e.g.\ Bailer-Jones 2015). This does not matter here because (a) we are only interested in nearby, bright stars (within 20\,pc), for which the uncertainties are very small, and (b) we are only using Hipparcos to give a realistic model for the distribution of real stars; we are not yet trying to navigate a real spacecraft, for which we would of course want highly accurate data.
The Gaia survey (Gaia Collaboration 2016) is currently providing much more accurate data than Hipparcos, but as Gaia has not yet published data on some of the nearest stars, I use Hipparcos.  In reality we would also want to remove known binary stars, as their space motions over tens to hundreds of years could deviate significantly from the linear motion assumed in section~\ref{sec:star_motions}. 

\begin{figure}
\begin{center}
\includegraphics[width=0.5\textwidth, angle=0]{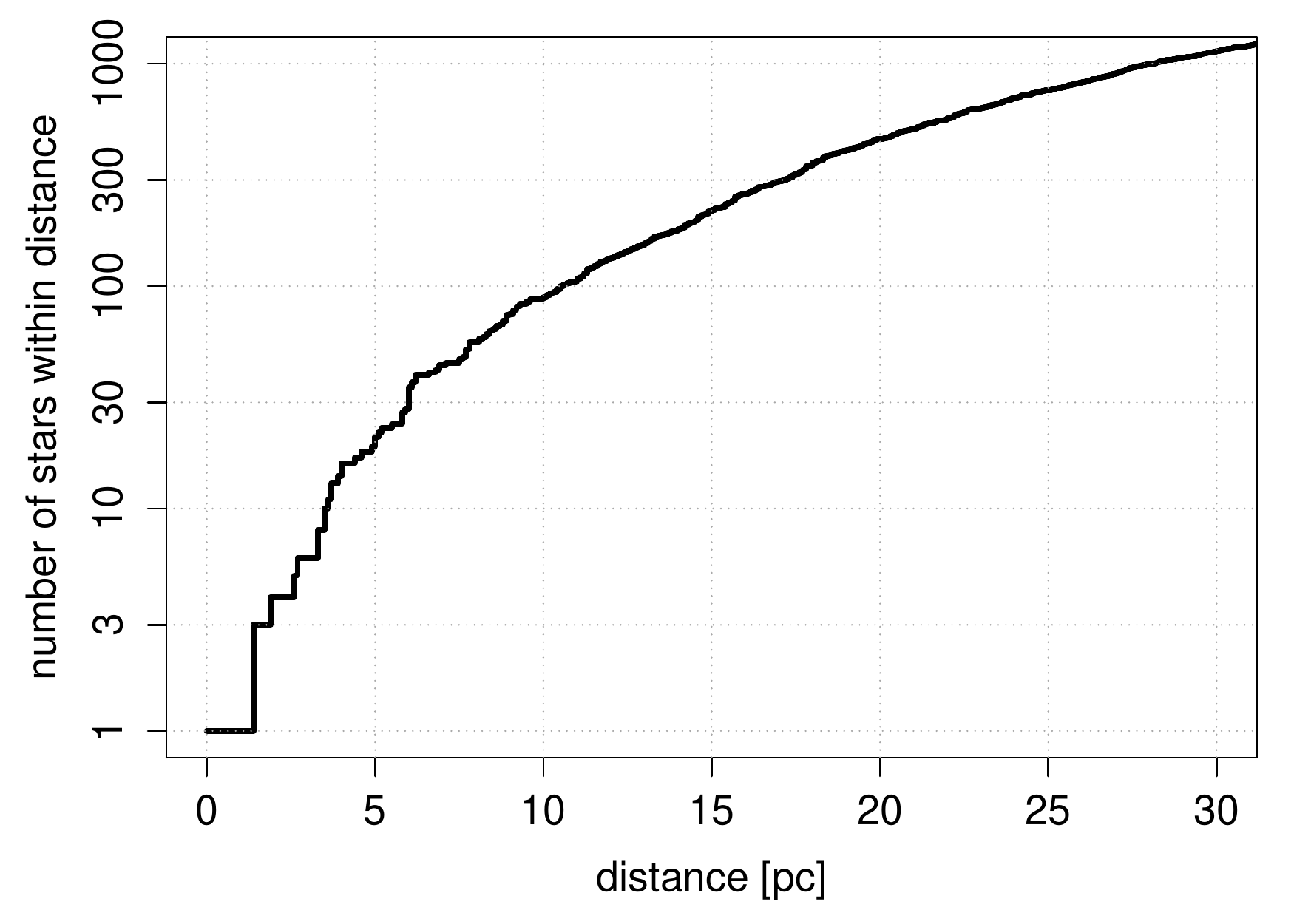}
\caption{The number of stars within a given distance in the star catalogue (the closest star is the Sun). Where stars are in binary systems the line jumps by two stars at a given distance (e.g.\ $\alpha$~Centauri A and B being the second and third closest stars). 
\label{fig:star_catalogue_distances}
}
\end{center}
\end{figure}

\begin{figure}
\begin{center}
  \includegraphics[width=0.6\textwidth, angle=0]{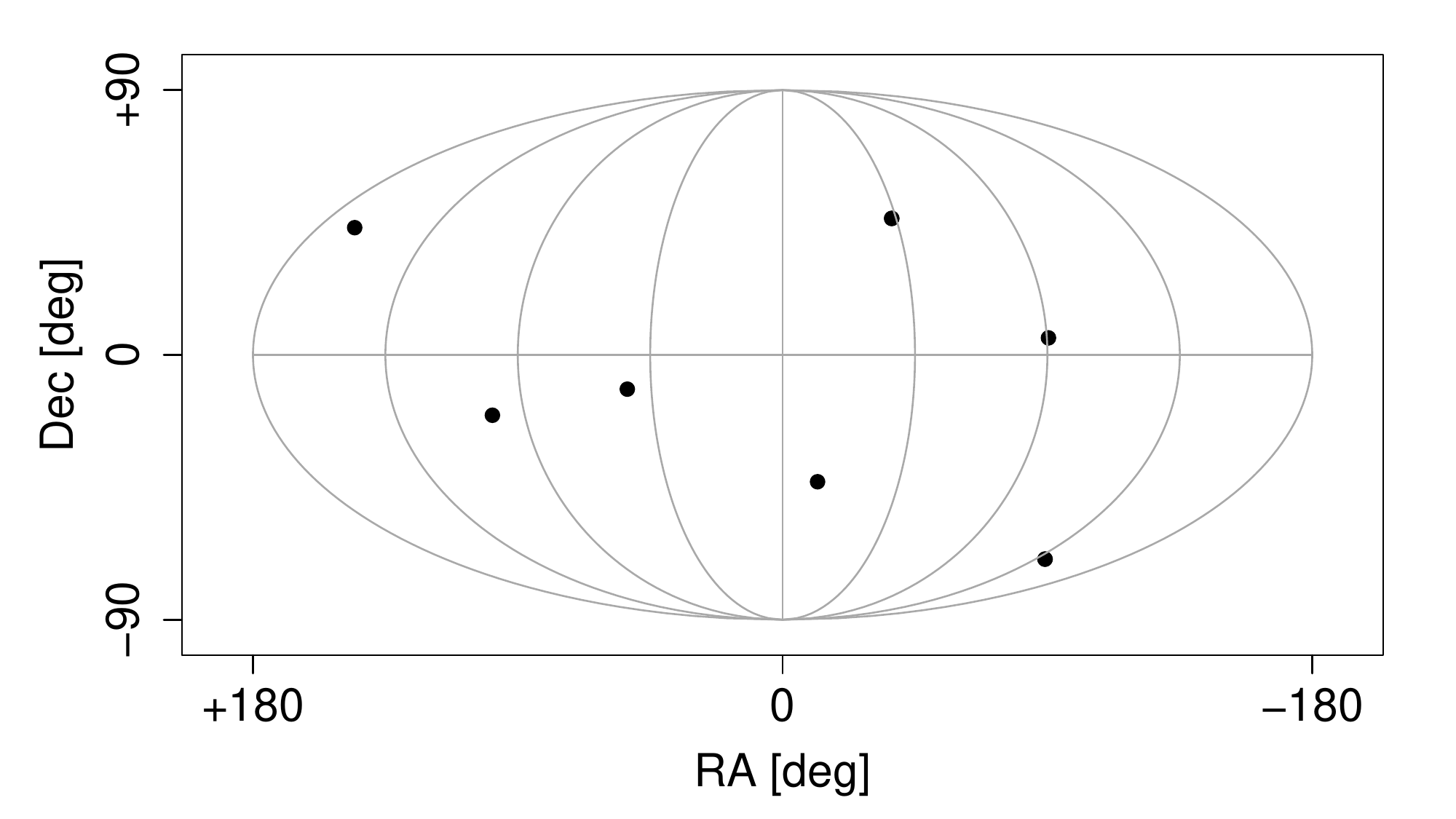}
  \includegraphics[width=0.6\textwidth, angle=0]{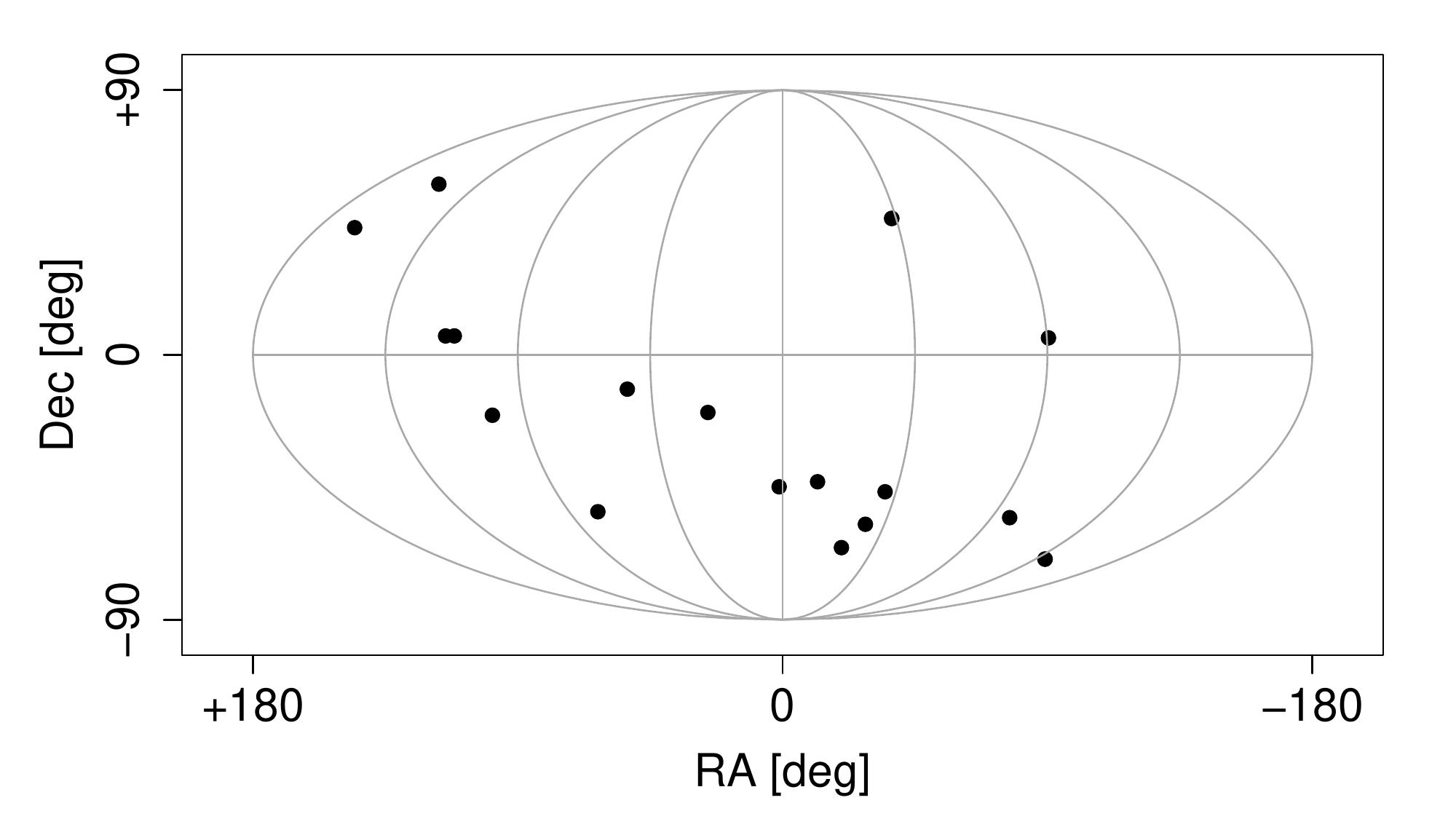}
  \includegraphics[width=0.6\textwidth, angle=0]{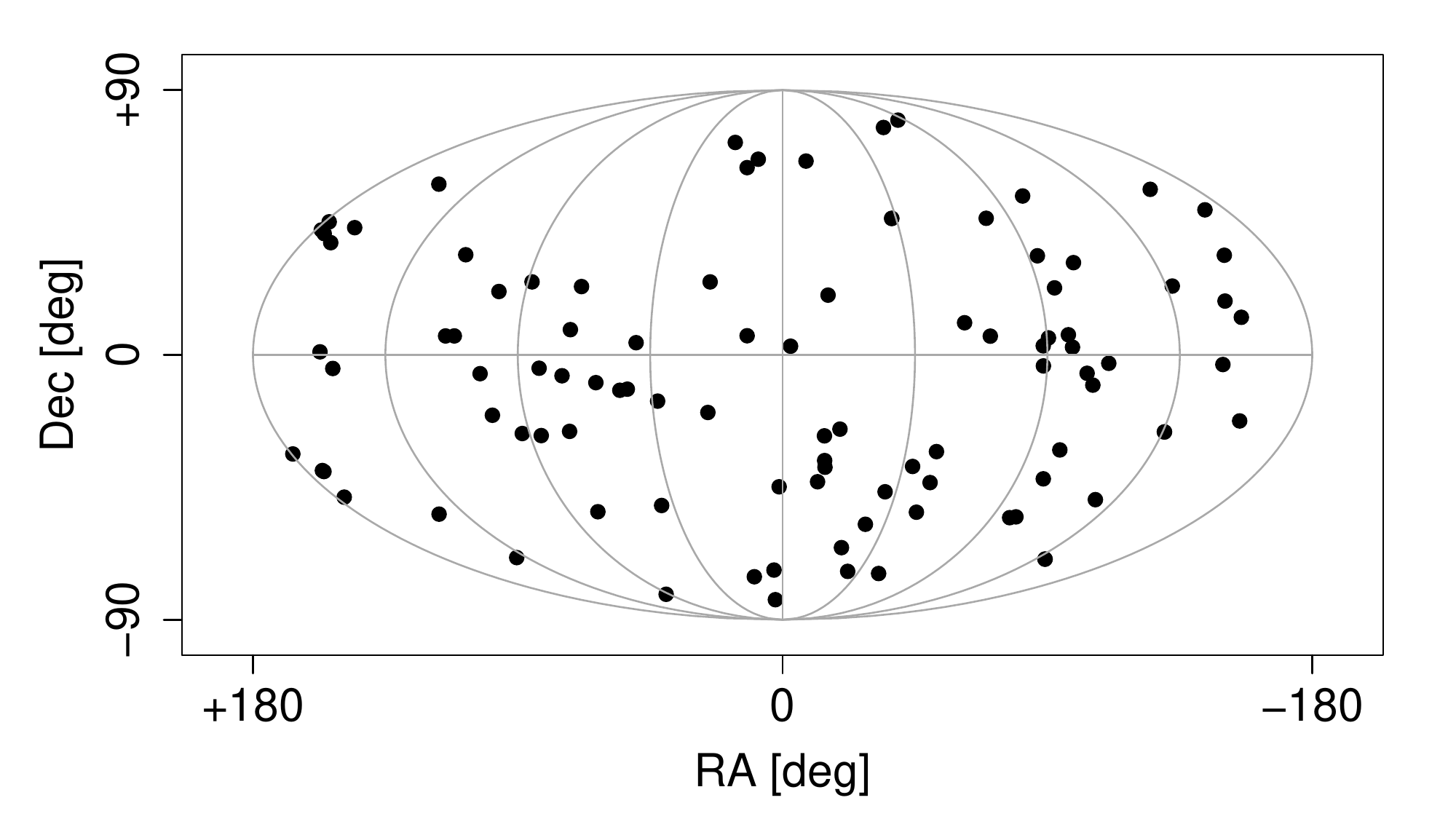}
\caption{Distribution of the nearest 10, 20, and 100 stars (top to bottom) on the sky in equatorial coordinates (the Sun is not shown) in the SSB reference frame (the ICRS). Some stars are in binary systems so appear as a single point. The projection is equal-area Mollweide.
\label{fig:star_catalogue_skymap}
}
\end{center}
\end{figure}
 
The first generation interstellar spacecraft are likely to be small, so they will be limited to small navigation systems that can only observe brighter stars. I therefore restrict the star catalogue just to those stars brighter than a visual magnitude of 10.  The number of bright stars within different distances is shown in Figure~\ref{fig:star_catalogue_distances}. The sky locations of the nearest 10, 20, and 100 stars are shown in Figure~\ref{fig:star_catalogue_skymap}.
% The smallest parallax displacement our spacecraft will have to measure is when the spacecraft moves the least and observes the most distant star. Taking these as 0.01\,\pc\ and 20\,\pc\ respectively, this gives a parallax of 100\,\as

The choice of reference star for the angular distance measurements is arbitrary, although in practice it should be a bright stable source. I will use the Sun.
%%% For convenience it is actually the SSB. This is a tiny difference, and one that we could anyway accommodate with no impact. 
In the simulations in section~\ref{sec:results} we will look at how the performance varies with the number $N$ of stars used (of which one is the Sun).  As nearer stars are better for navigation, due to their larger parallaxes, whenever I select $N$ stars, this is always the $N-1$ nearest stars to the Sun plus the Sun itself.

\subsection{True spacecraft parameters}\label{sec:truepars}

In section~\ref{sec:results} we will examine the performance of the method in various scenarios, such as differing number of stars used.  For each scenario I place the spacecraft at 100 different randomly-drawn positions, velocities, and times, infer the spacecraft parameters for each, and then average the performance over these 100 runs (exact details of the performance metrics are given in section~\ref{sec:metrics}). This characterizes the performance of the scenario more robustly than just a single run.

To generate the true position of the spacecraft, I draw a random position on the sky and a random distance uniformly between 0.1 and 10\,\ly\ (0.31--3.1\,\pc), the upper limit being characteristic of how far the first generation interstellar spacecraft are likely to reach.
As the spacecraft will spend most of its mission moving directly away from the Sun, I set the true velocity vector, \xvecdot,  parallel to the position vector, \xvec, but draw the magnitude of the velocity from a uniform distribution.\footnote{Parallel position and velocity vectors is not a necessary restriction, just a realistic one. The performance of the inference is not altered significantly when relaxing this.}
Here I distinguish between two cases:
relativistic, where the magnitude of the velocity can range from 0 to 0.5\,c, and
non-relativistic, where the magnitude of the velocity can range from 0 to 500\,\kms.
The true SSB time of the measurement, \timeobs, is drawn from a uniform distribution between 10 and 20\,\yr. Note that there is no need for the distance, velocity, and time to be self-consistent, 
because \timeobs\ is the time of the observation, not the duration of the spacecraft's journey.
%The spacecraft may anyway have moved with variable speed up to the point of measurement.
The random selection of true spacecraft parameters is done anew for every scenario.

\subsection{Performance metrics}\label{sec:metrics}

I take the median of the MCMC chains as the estimate of the spacecraft 3D position, 3D velocity, and the time. For each parameter I compute the residual, the difference between the estimate and the true value. For the position and velocity I then compute the magnitude of the residuals, i.e.\ $|\xvec_{\rm inferred} - \xvec_{\rm true}|$ and $|\xvecdot_{\rm inferred} - \xvecdot_{\rm true}|$.
%For the time I compute $|t_{\rm inferred} - t_{\rm true}|$. 
These give us how far away we are in position, velocity, and (modulus of) time from the true values. These three positive quantities I refer to as the {\em accuracies}.  The uncertainties in the seven model parameters are obtained from the standard deviations of the MCMC samples.  I compute the uncertainty in the magnitude of the position residual via a first order propagation of the uncertainties in its three components, taking into account the correlations between the three components (computed from the MCMC samples). I do the same for the uncertainty in the velocity.  I refer to these uncertainty estimates in the position, velocity, and (modulus of) the time as the {\em precisions}.
% All of these performance metrics are quantities in the SSB rest frame.

As each scenario involves multiple instantiations of the true spacecraft parameters, I compute the median accuracy and median precision over these runs to give the accuracy and precision for that scenario.  To get some idea of the spread in the accuracy over the runs, I also compute the lower and upper 1-$\sigma$ quantiles over the set of runs in that scenario.

%%%%%%%%%%%%%%%%%%%%%%%%%%%%%%%%%%%%%%%%%%%%%%%%%%%%%%%%%%%%%%%%%%%%%%%
\section{Simulation results: navigation accuracy and precision}\label{sec:results}

%%% Nominal_simulation is the same set up as inf_7par_Ntruepar20_rel_1arcsec_10kms_both_Nstar_10
%%% but is a different initial random seed. Was re-run so I could produce new MCMC plots, which are not saved to the .Robj.

\subsection{Nominal scenario}

We start by looking at the performance in the nominal scenario, which has the following properties: spacecraft moving at relativistic velocities (as defined in section~\ref{sec:truepars}); 1\as\ angular distance measurement accuracy; 10\,\kms\ radial velocity measurement accuracy; $N=10$ stars.  This angular accuracy is similar to what can be achieved by commercial cm-sized star trackers.
%%% N08
The radial velocity accuracy can be achieved with a spectrograph of spectral resolution a few thousand.

\begin{figure}
\begin{center}
\includegraphics[width=1.0\textwidth, angle=0]{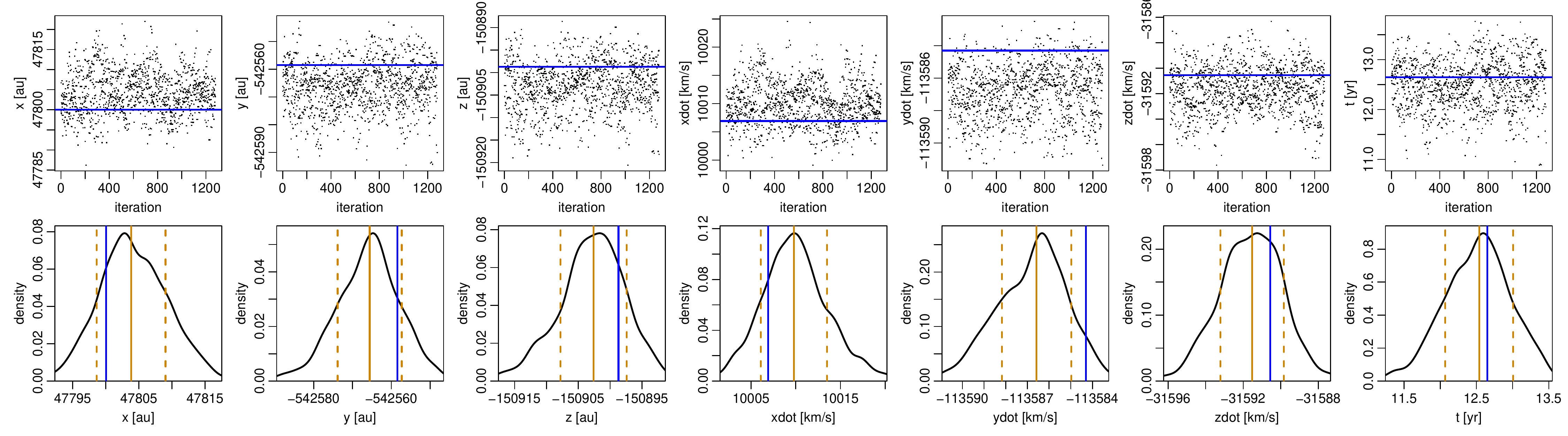}
\caption{MCMC chain (top) and probability density function of these same samples (bottom) for one run of the nominal scenario
(1\as\ angular distance measurement accuracy, 10\,\kms\ radial velocity measurement accuracy, $N=10$ stars).
The horizontal blue lines in the upper panels and the vertical blue lines in the lower panels show the true parameters. The solid orange lines in the lower panels are the median of the samples, which we take to be our parameter estimates. The dashed vertical orange lines show $\pm1$ standard deviation around the median. $x, y, z$ are Cartesian coordinates in the ICRS.
\label{fig:inf_7par_Ntruepar20_rel_1arcsec_10kms_both_Nstar_10_postSamp_1}
}
\end{center}
\end{figure}

\begin{figure}
\begin{center}
\includegraphics[width=1.0\textwidth, angle=0]{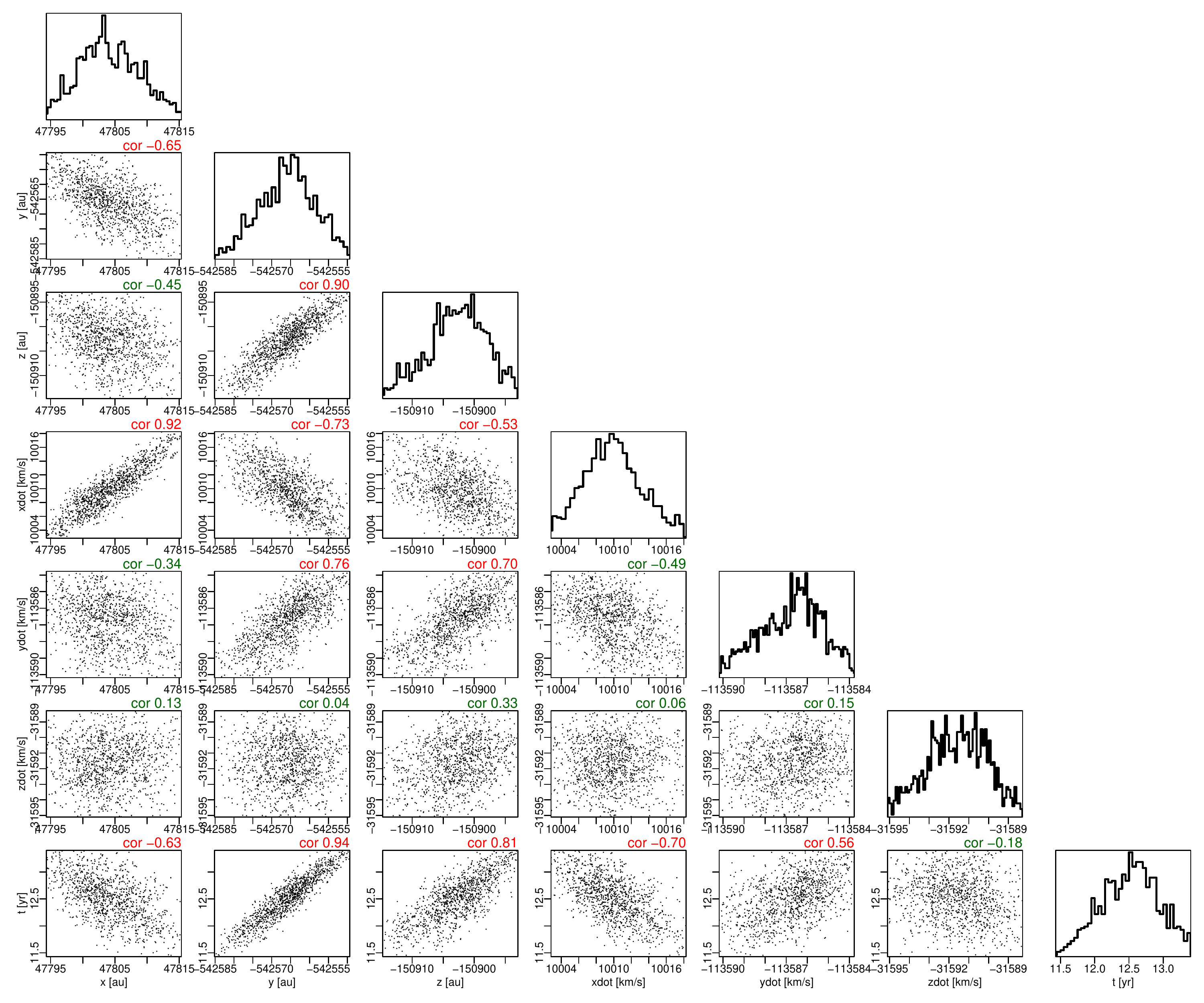}
\caption{Pairwise correlations between the MCMC samples for the run shown in Figure~\ref{fig:inf_7par_Ntruepar20_rel_1arcsec_10kms_both_Nstar_10_postSamp_1}. The number at the top-right of each panel is the correlation coefficient.
\label{fig:inf_7par_Ntruepar20_rel_1arcsec_10kms_both_Nstar_10_postSamp_cornerplot_1}
}
\end{center}
\end{figure}

We look first at one run of the nominal scenario in which the spacecraft is 8.9\,\ly\ from the SSB and travelling at 0.39\,c.  The samples from the MCMC are shown in Figure~\ref{fig:inf_7par_Ntruepar20_rel_1arcsec_10kms_both_Nstar_10_postSamp_1}.
The burn-in has been discarded and the sampling has entered a steady state.  The pair-wise correlations of the samples are shown in Figure~\ref{fig:inf_7par_Ntruepar20_rel_1arcsec_10kms_both_Nstar_10_postSamp_cornerplot_1}. Some parameters are quite correlated. The degree of correlation varies according to the true spacecraft parameters.

\begin{figure}
\begin{center}
\includegraphics[width=1.0\textwidth, angle=0]{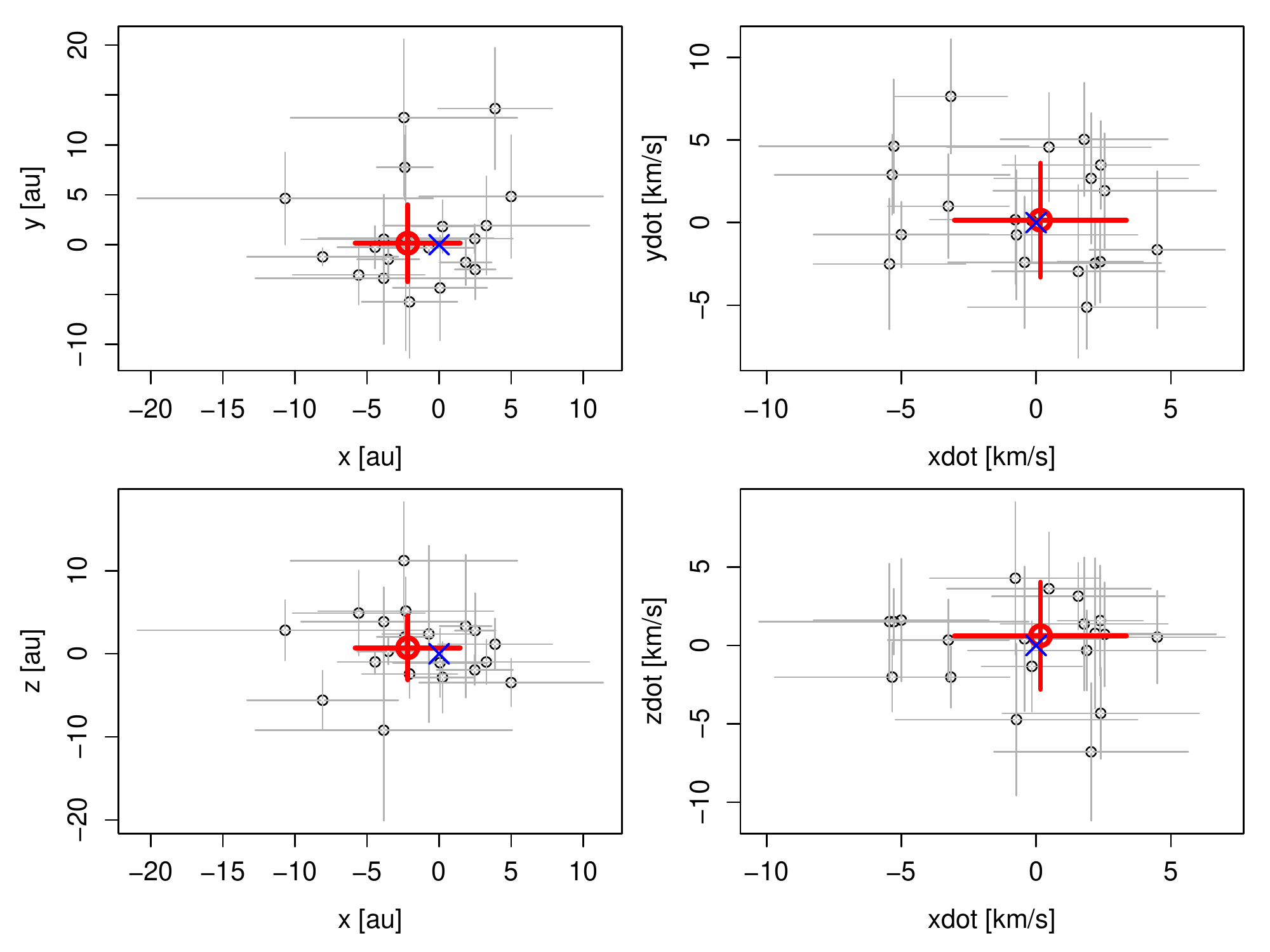}
\caption{The open black circles show the residuals (estimated minus true) for the three position and three velocity parameters (different panels) for 20 runs of the nominal scenario
(1\as\ angular distance measurement accuracy, 10\,\kms\ radial velocity measurement accuracy, $N=10$ stars).
The blue cross indicates zero residual. The (symmetric) black error bars are the standard deviation of the MCMC samples for each run, so give some idea of the uncertainty of the estimate. The open red circle is the median residual of all the individual runs, so gives some idea of estimation bias (this is not the accuracy). The thick red error bar shows the median uncertainty.
\label{fig:inf_7par_Ntruepar20_rel_1arcsec_10kms_both_Nstar_10_resComp}
}
\end{center}
\end{figure}

We now repeat this simulation for 100 runs, i.e.\ 100 different random instantiations of the true spacecraft parameters (see section~\ref{sec:truepars}). The residuals -- the differences between the estimates and their true values -- for the spacecraft position and velocity coordinates for 20 of these runs are shown in Figure~\ref{fig:inf_7par_Ntruepar20_rel_1arcsec_10kms_both_Nstar_10_resComp} as open black circles.  These show a scatter about the true coordinates (blue cross), as expected.
% Some of this is scatter due to the random noise in the measurement, and some due to the different spacecraft coordinates.
The uncertainties in the estimated coordinates, given by the black error bars, are similar in size to this scatter, showing that the MCMC sampling is capturing the irreducible noise in the data. The red open circle in Figure~\ref{fig:inf_7par_Ntruepar20_rel_1arcsec_10kms_both_Nstar_10_resComp} is the median over the runs.  If the inference is unbiased, then the inferred components should be as often above as below the true values, in which case we expect this median residual to be near to zero, which is what we see.
%%% To quantify, it would be need to be near zero to within the uncertainties (given by the red error bars)
%%% divided by sqrt(N). But this is going into too much detail, plus we only show 20 runs in this example.

\begin{figure}
\begin{center}
\includegraphics[width=1.0\textwidth, angle=0]{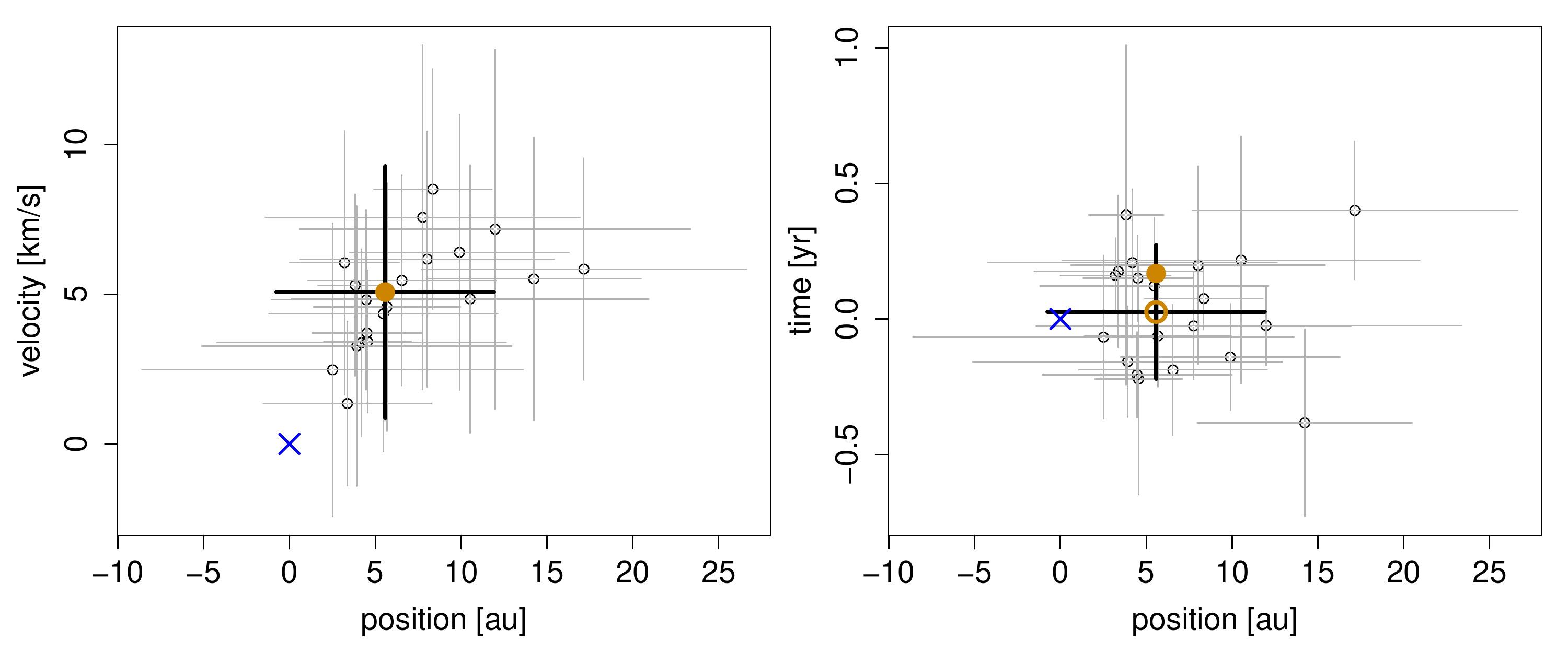}
\caption{Left: The open black circles show the magnitude of the position and velocity residuals for 20 runs, computed from their component residuals shown in Figure~\ref{fig:inf_7par_Ntruepar20_rel_1arcsec_10kms_both_Nstar_10_resComp}. These magnitudes are of course always positive.
The thin black error bars are computed by propagating the uncertainties in the components taking into account the correlations.
%%% The error bars can extend to negative values because they are computed from the standard deviation.
%%% We can't change to using asymmetric quantiles, because we couldn't then do (simple) error propagation
%%% from the component quantiles.
The filled orange circle is the median of the black open circles, and the thick black error bar is the median of the thin black error bars. These are the accuracy and precision (respectively) of our inference. Right: The horizontal axis is the same as in the left panel; the vertical axis is the time residual. The open orange circle is the median of the time residual; as this has a sign it is a measure of the bias over the 20 runs. The filled orange circle is the median of the {\em absolute} time residual, and so measures the accuracy. The blue cross shows zero residual.
\label{fig:inf_7par_Ntruepar20_rel_1arcsec_10kms_both_Nstar_10_resMag}
}
\end{center}
\end{figure}

From the residuals in the three position components we can compute the {\em magnitude} of the position residual, and likewise for the velocity.  These are plotted in the left panel of Figure~\ref{fig:inf_7par_Ntruepar20_rel_1arcsec_10kms_both_Nstar_10_resMag} as the open black circles. These are the overall {\em accuracies} with which we have determined the position and velocity. We see that they are broadly consistent with their corresponding uncertainties, the {\em precisions}, shown as thin black error bars, which are computed by propagating the component uncertainties (taking into account the correlations, e.g.\ Figure~\ref{fig:inf_7par_Ntruepar20_rel_1arcsec_10kms_both_Nstar_10_postSamp_cornerplot_1}).  Hence the precisions, which we can obtain from the MCMC even when we don't know the true spacecraft coordinates, are a useful estimate of the accuracies.  The median accuracy over the runs is shown by the filled orange circle, and the median precision by the thick black error bar.  This shows that in the nominal scenario we can determine the position of the spacecraft to within about 5\,\au\ and the velocity to within about 5\,\kms.

The right panel of Figure~\ref{fig:inf_7par_Ntruepar20_rel_1arcsec_10kms_both_Nstar_10_resMag} shows the (signed) time residuals. The median of the magnitude of these is again shown as an filled orange circle.  Time is less well determined, to only about 0.2\,yr. This is not surprising, because time only enters the inference via the changes in the stars' positions, and these are small compared to the distances to the stars and spacecraft (see equation~\ref{eqn:svecprime}).  It is also unimportant, because time is not one of the parameters we are actually interested in.  Indeed, if we artificially set the stars' velocities to zero, then we find that time is entirely unconstrained, as expected.
% See results/nominal_scenario_zero_PM/
The accuracy with which the other parameters can be inferred remains unaffected.

\begin{figure}
\begin{center}
\includegraphics[width=1.0\textwidth, angle=0]{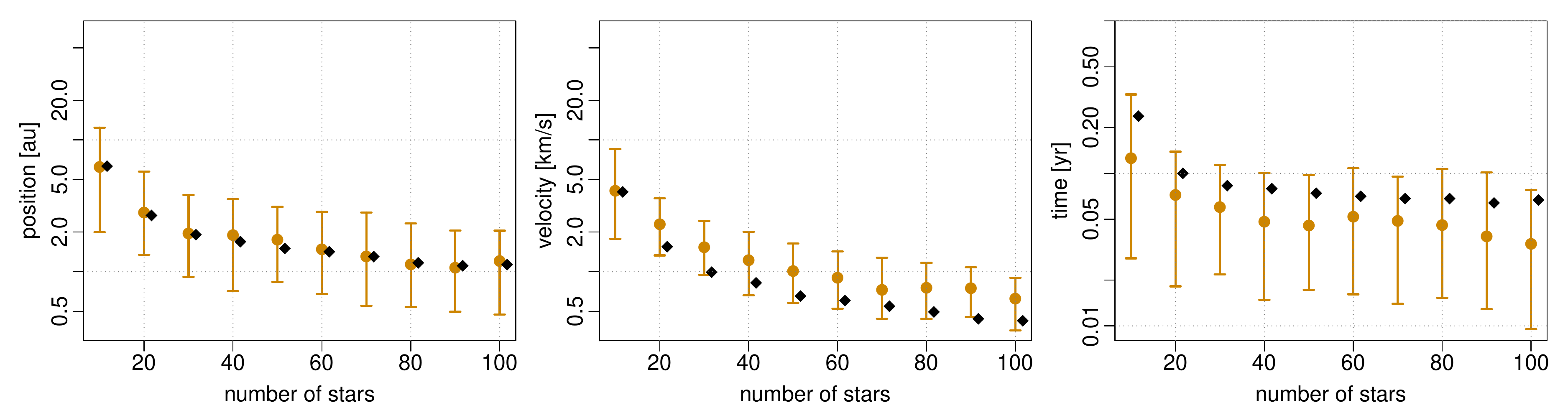}
\caption{Median accuracy (filled orange circles) and median precision (black diamonds) over 100 runs for simulations using different number of stars. The other parameters are as in the nominal scenario: spacecraft moving at relativistic velocities; 1\as\ angular distance and 10\,\kms\ radial velocity measurement accuracies.
These statistics are for the {\em magnitudes} of the spacecraft parameters: the accuracy in each of the three spatial or velocity components is on average $\sqrt{3}$ times smaller.
  The orange error bars show the spread in accuracies over the 100 runs, the lower bar being the 16th percentile and the upper bar the 84th percentile, to give asymmetric 1$\sigma$-like error bars. 
  The black diamonds are slightly offset in the horizontal direction for clarity.
\label{fig:inf_7par_Ntruepar100_resAll_rel_1arcsec_10kms_both}
}
\end{center}
\end{figure}

\subsection{Variations of the number of stars and the measurement accuracies}

In this section we will see how the performance varies as we change the number of stars used, the accuracy of the measurements, and whether we use only angular positions or only radial velocities.

We first repeat the nominal scenario but for different numbers of stars, ranging from 10 to 100 in steps of 10. The median accuracy and precision averaged over the 100 runs in each case are shown in Figure~\ref{fig:inf_7par_Ntruepar100_resAll_rel_1arcsec_10kms_both}
as the orange circles and black diamonds respectively.
As expected, both metrics improve as we use more stars.
For example, with 100 stars at the nominal measurement accuracy we can locate our spacecraft in deep space to within 1.2\,\au\ and determine its velocity to better than 0.6\,\kms.
The orange error bars show the (asymmetric) 1$\sigma$-like range on the accuracies across the runs. Thus while the median position accuracy using, for example, 20 stars is 2.8\,\au, the (asymmetric) 1$\sigma$ range is 1.3--5.8\,\au.
We see in Figure~\ref{fig:inf_7par_Ntruepar100_resAll_rel_1arcsec_10kms_both} that the 
estimated precisions (black diamonds) reflect the true accuracies (orange circles) very well for the spacecraft position (left panel).
For the velocity, the precisions slightly underestimate the accuracies, i.e.\ are optimistic (central panel), whereas the opposite is true for the time (right panel).

\begin{figure}
\begin{center}
\includegraphics[width=1.0\textwidth, angle=0]{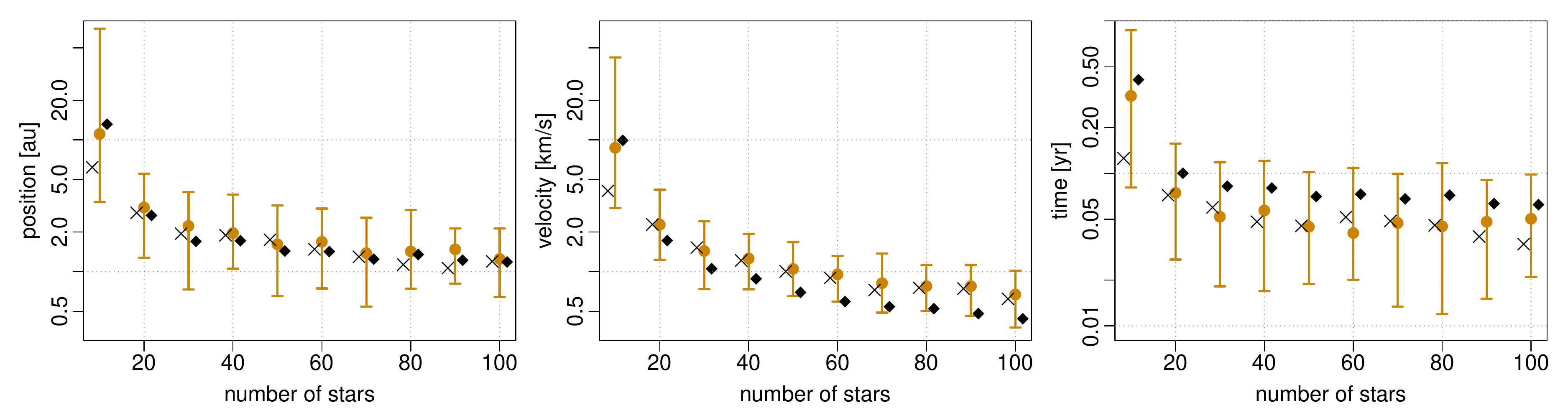}
\caption{As Figure~\ref{fig:inf_7par_Ntruepar100_resAll_rel_1arcsec_10kms_both}, but for a scenario in which we only use angular distance measurements.
The black crosses (slightly offset in the horizontal direction for clarity) show the median accuracies from Figure~\ref{fig:inf_7par_Ntruepar100_resAll_rel_1arcsec_10kms_both}.
\label{fig:inf_7par_Ntruepar100_resAll_rel_1arcsec_angdist}
}
\end{center}
\end{figure}

\begin{figure}
\begin{center}
\includegraphics[width=1.0\textwidth, angle=0]{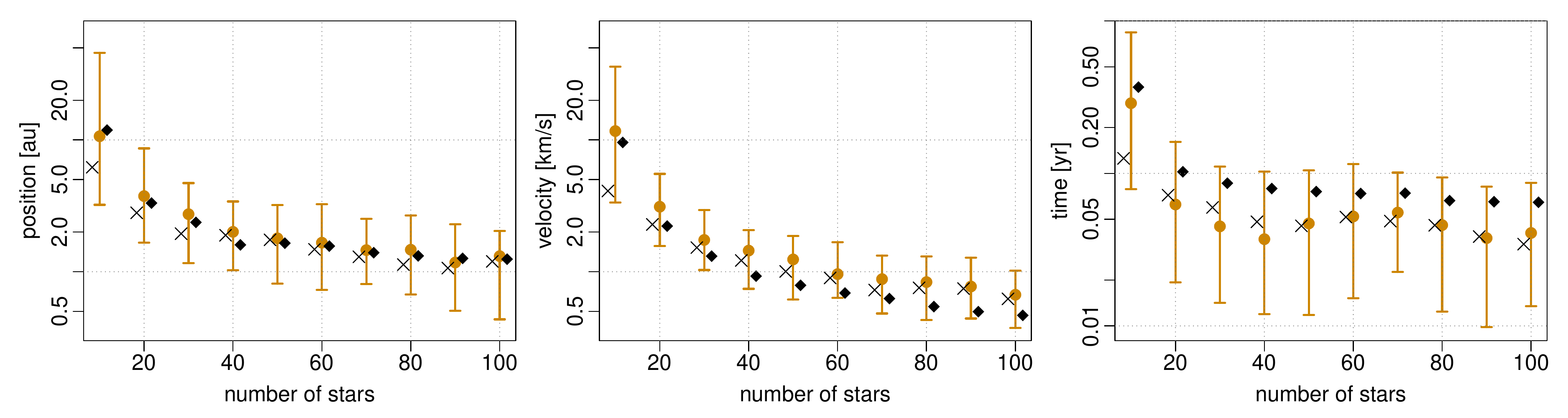}
\caption{As Figure~\ref{fig:inf_7par_Ntruepar100_resAll_rel_1arcsec_angdist}, but for a scenario in which the spacecraft is moving non-relativistically (0 to 500\,\kms\ as opposed to 0 to 150\,000\,\kms).
\label{fig:inf_7par_Ntruepar100_resAll_nonrel_1arcsec_angdist}
}
\end{center}
\end{figure}

Figure~\ref{fig:inf_7par_Ntruepar100_resAll_rel_1arcsec_angdist} shows the same scenarios just investigated, but now using only angular distance measurements, i.e.\ no radial velocities. To ease comparison, the black crosses show the median accuracies for previous case.  The performances are very similar. Thus radial velocities of this degree of accuracy do not improve the navigational accuracy, and the spacecraft velocity can be inferred using just the aberration of the stars' positions. Aberration is of course large when moving at relativistic velocities, so we might expect worse performance at non-relativistic speeds (0--500\,\kms) when only using angular distance measurements.  This is not the case, however, as we see in Figure~\ref{fig:inf_7par_Ntruepar100_resAll_nonrel_1arcsec_angdist}: the performance is just as good.  The performance also changes little if we include radial velocity measurements again (at the nominal accuracy of 10\,\kms; plot not shown).  The accuracy with which we can determine the velocity of the spacecraft is therefore independent of its velocity in this accuracy regime.

\begin{figure}[t]
\begin{center}
\includegraphics[width=1.0\textwidth, angle=0]{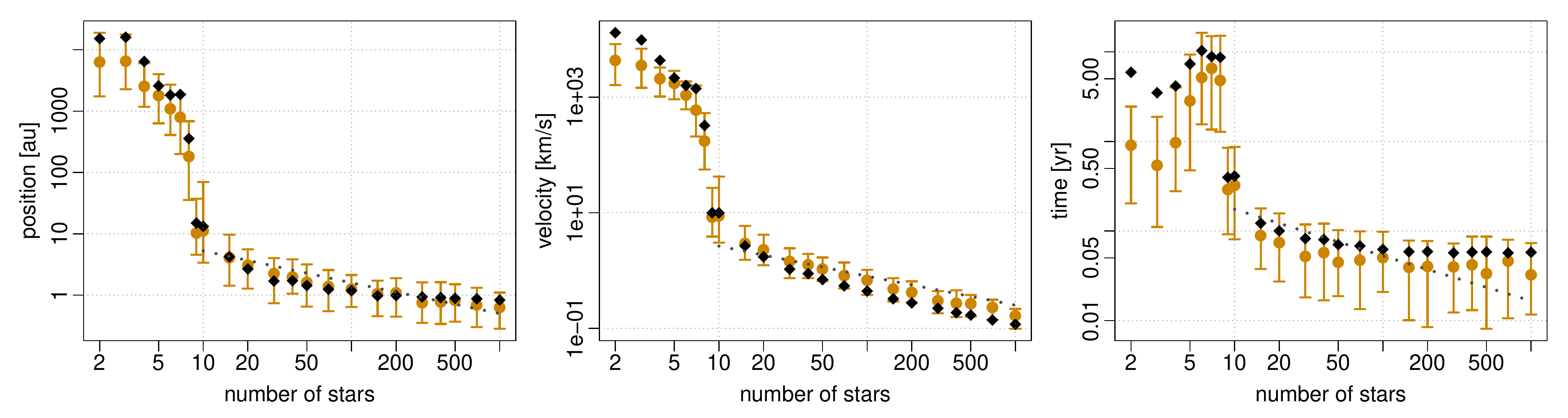}
\caption{Performance for a scenario in which the spacecraft is moving relativistically and we only use angular distance measurements (at 1\as\ precision).
This is similar to Figure~\ref{fig:inf_7par_Ntruepar100_resAll_rel_1arcsec_angdist} but now for a larger range of the number of stars, and logarithmic scales on the horizontal axes. The diagonal dotted grey lines show a power law $(N-1)^{-1/2}$ with arbitrary vertical offset. The accuracy variation would be parallel to this if it varied in a central limit theorem-like way.
\label{fig:inf_7par_Ntruepar100_resAll_rel_1arcsec_angdist_NstarLargeRange}
}
\end{center}
\end{figure}

Returning to the relativistic case, we ask ourselves how well we can navigate with a much smaller or larger number of stars, $N$, just using angular distance measurements. The minimum number of stars is two, i.e.\ one measurement, although as we need to solve for seven model parameters we expect poor performance. This is confirmed in Figure~\ref{fig:inf_7par_Ntruepar100_resAll_rel_1arcsec_angdist_NstarLargeRange}.
For seven or fewer stars we see very poor MCMC chains that move little from the initialization.
%%% They get gradually worse from 7 down to 2. At 2 there is very little updating.
The accuracy of position and velocity determination improve as we add more stars.\footnote{$\alpha$~Centauri A and B have identical parallaxes in the catalogue. It so happens that B is taken as the second closest and A as the third closest. Even though they also have almost the same position on the sky, their velocities differ, so there is still an improvement in going from two to three stars with this catalogue.}
We see a sharp increase in performance from 8 to 9 stars, i.e.\ 7 to 8 measurements, which is the transition from formally having non-degenerate equations to a redundancy of one. As we increase the number of stars further we see a slower but steady improvement in position and velocity accuracy.  If performance improved according to the central limit theorem, we would expect the accuracy to vary as $(N-1)^{-1/2}$.  Such a variation is shown by the diagonal dotted grey lines in Figure~\ref{fig:inf_7par_Ntruepar100_resAll_rel_1arcsec_angdist_NstarLargeRange}. While position improves roughly at this rate, the velocity improves slightly more rapidly.  The variation of the time accuracy is much more erratic. It generally improves up to a few tens of stars, but then appears to plateau. Note that as we increase the number of stars we also increase the average distance to the stars (Figure~\ref{fig:star_catalogue_distances}), and this itself may (negatively) impact the performance (see section~\ref{sec:other_variations}).

\begin{figure}
\begin{center}
\includegraphics[width=1.0\textwidth, angle=0]{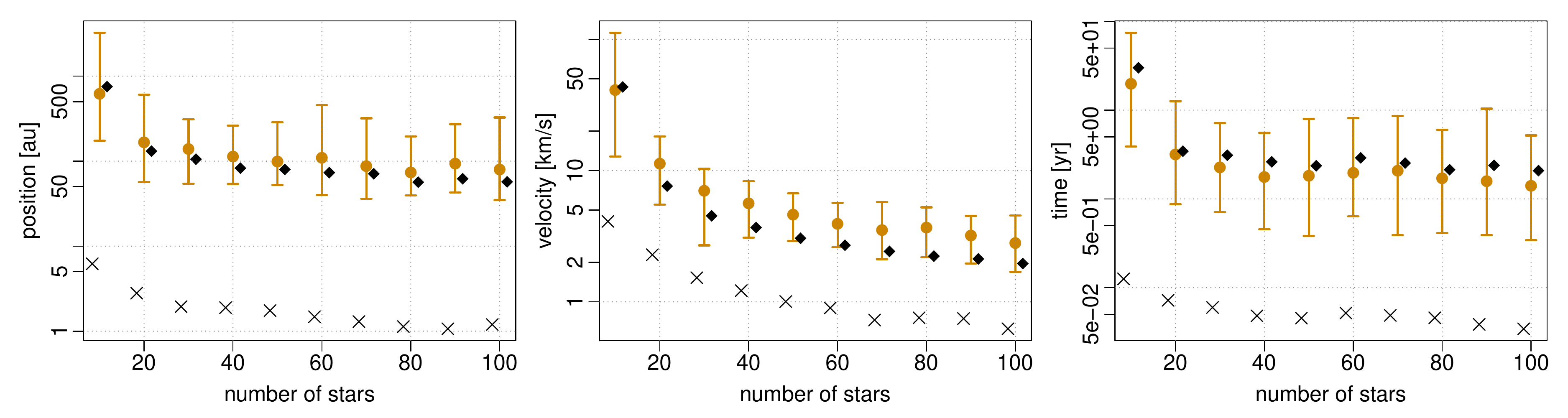}
\caption{As Figure~\ref{fig:inf_7par_Ntruepar100_resAll_rel_1arcsec_10kms_both}, but for a scenario in which we only use the radial velocity measurements (at 10\,\kms\ precision). Note the change in the ranges of the vertical axes.
\label{fig:inf_7par_Ntruepar100_resAll_rel_10kms_rv}
}
\end{center}
\end{figure}

Can we navigate using only radial velocities? The performance when adopting a 10\,\kms\ measurement accuracy is shown in Figure~\ref{fig:inf_7par_Ntruepar100_resAll_rel_10kms_rv}: It is considerably worse than when we use also the angular distances measurements (Figure~\ref{fig:inf_7par_Ntruepar100_resAll_rel_1arcsec_10kms_both}) or indeed only these (Figure~\ref{fig:inf_7par_Ntruepar100_resAll_rel_1arcsec_angdist}).

\begin{figure}
\begin{center}
  \includegraphics[width=1.0\textwidth,angle=0]{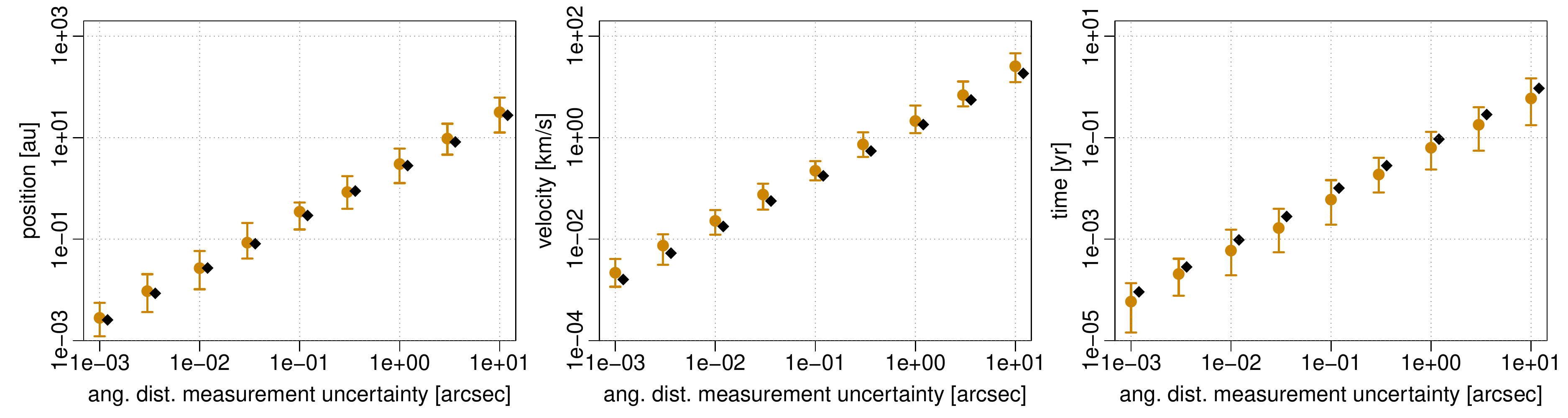}
\caption{Variation of performance with the accuracy of the angular distance measurements when using 20 stars. Radial velocity measurements are not used. The spacecraft is moving relativistically. The plotted quantities are as described in 
 Figure~\ref{fig:inf_7par_Ntruepar100_resAll_rel_1arcsec_angdist}.
\label{fig:resAll_rel_Nstar20_angdist_angdistNoise}
}
\end{center}
\end{figure}

Let us now examine how the performance varies with measurement accuracy. For this we will use 20 stars, as in the nominal scenario.  First we use only angular distance measurements. The performance variation is shown in Figure~\ref{fig:resAll_rel_Nstar20_angdist_angdistNoise}. We see a near-linear variation on what is a log-log plot, meaning there is probably a power-law relation between the accuracy of the inference (vertical axis, $y$) and the accuracy of the measurements (horizontal axis, $x$) of the form $y = ax^b$. If we fit a linear model to the logged quantities
% $\log y = \log a + b\log x$,
then we find that $b=1$ to within the uncertainties for all three quantities (position, velocity, time). This is plausible: changing the measurement accuracy by some factor changes the achieved accuracy by the same factor.
We learn from Figure~\ref{fig:resAll_rel_Nstar20_angdist_angdistNoise} that if we could measure angular distances to 0.1\as\ with 20 stars, then we could determine the location of our spacecraft to 0.3\,\au\ and its velocity to 0.2\,\kms.
To achieve this same positional accuracy with a measurement accuracy of 1\as\ -- 10 times worse -- 
we would need 100 times as many stars, i.e.\ 2000, assuming that the approximate $1/\sqrt{N}$ performance scaling holds.
We saw in Figure~\ref{fig:inf_7par_Ntruepar100_resAll_rel_1arcsec_angdist_NstarLargeRange} that the velocity accuracy improves slightly faster than $1/\sqrt{N}$, but that figure shows we would still need of order 1000 stars to achieve 0.2\kms\ velocity accuracy with the inferior measurements.
In other words, navigational accuracy gains are made faster by improving measurement accuracy than by observing more stars.
What the optimal trade-off is between measurement accuracy and number of stars depends on practical details of the mission, such as the payload mass, size, and stability, as well as the time available for observations.

We see in Figure~\ref{fig:resAll_rel_Nstar20_angdist_angdistNoise} that the orange circles and black diamonds for a given measurement generally agree within the 1$\sigma$ range of the former. This means that the precisions are generally good estimates of the accuracies. This is important because in a real application we of course do not know the true spacecraft coordinates so cannot compute the accuracy.

\begin{figure}
\begin{center}
\includegraphics[width=1.0\textwidth,angle=0]{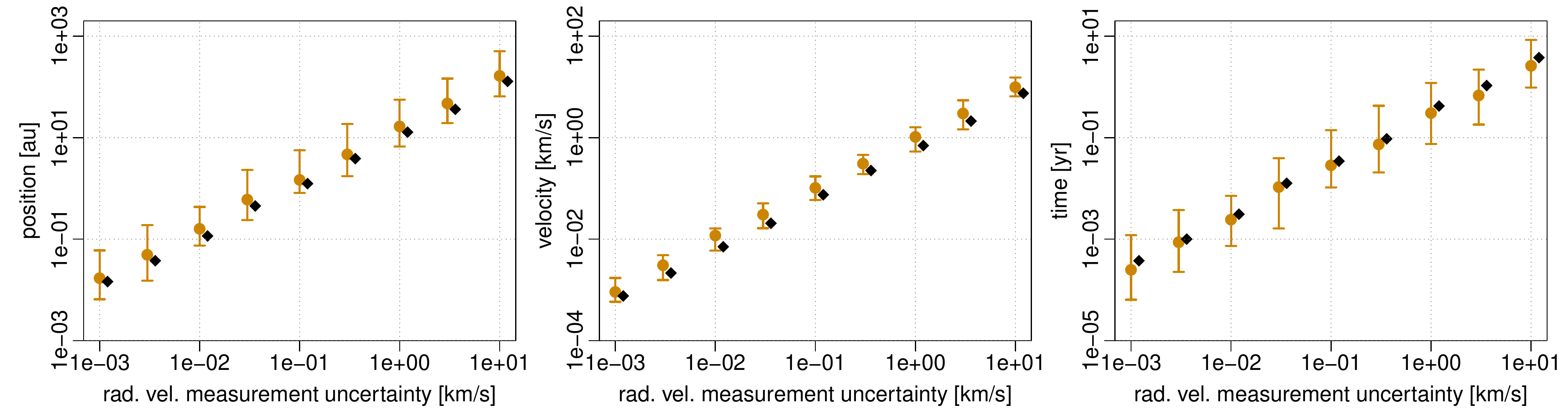}
\caption{Variation of performance with the accuracy of the radial velocity measurements when using 20 stars. Angular distance measurements are not used. The spacecraft is moving relativistically. The plotted quantities are as described in 
 Figure~\ref{fig:inf_7par_Ntruepar100_resAll_rel_1arcsec_angdist}.
\label{fig:resAll_rel_Nstar20_rv_rvNoise}
}
\end{center}
\end{figure}

We can analogously look at how the performance varies with radial velocity accuracy, when only using radial velocities. This is shown in Figure~\ref{fig:resAll_rel_Nstar20_rv_rvNoise} for 20 stars. We again see a power law variation of the parameter accuracy with the measurement accuracy, again with a power of 1.  It is rather unrealistic to think we could measure stellar radial velocities onboard our spacecraft to an accuracy of 0.01\,\kms, but if we could, we would be able to determine the position of our spacecraft to within 0.2\,\au\ on average using just the radial velocity measurements of 20 stars.

\begin{figure}
\begin{center} \includegraphics[width=1.0\textwidth,angle=0]{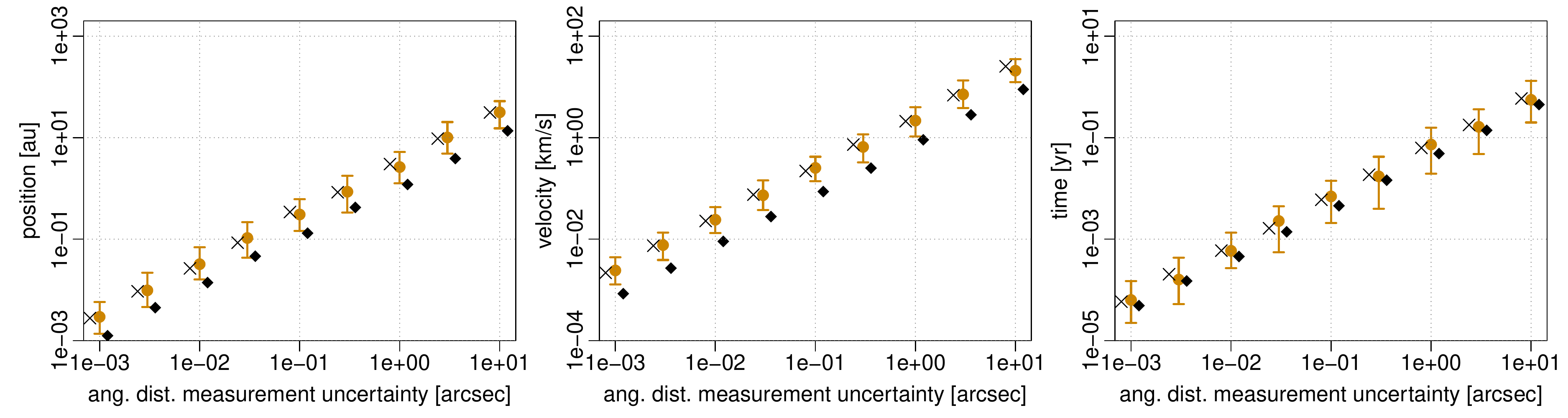}
\caption{As Figure~\ref{fig:resAll_rel_Nstar20_angdist_angdistNoise}, but now with the measurement accuracy in the likelihood ($\sigma_\angdist$\ in equation~\ref{eqn:likelihood_angdist}) underestimated by a factor of two. The black crosses
  (slightly offset in the horizontal direction for clarity) show the median accuracy from Figure~\ref{fig:resAll_rel_Nstar20_angdist_angdistNoise},
  i.e.\ when the measurement accuracy is correctly estimated.
\label{fig:resAll_rel_Nstar20_angdist_uncinfac0p5_angdistNoise}
}
\end{center}
\end{figure}

In all the simulations so far I have assumed that our estimates of the standard deviation of the measurement noise that we use in the likelihoods -- $\sigma_\angdist$\ in equation~\ref{eqn:likelihood_angdist} and $\sigma_\rv$\ in equation~\ref{eqn:likelihood_rv} -- are equal to the true standard deviations of the noise, i.e.\ to what we use to generate the noisy data in equations~\ref{eqn:noise_angdist} and~\ref{eqn:noise_rv}.  With a good understanding of the noise sources this should be achievable, but sometimes our estimate of the noise is systematically wrong. To investigate the impact of this, I redo the set of simulations for the variable angular distance measurement accuracy (those in Figure~\ref{fig:resAll_rel_Nstar20_angdist_angdistNoise}), but now underestimating the standard deviation used in the likelihood by a factor of two.  The results are shown in Figure~\ref{fig:resAll_rel_Nstar20_angdist_uncinfac0p5_angdistNoise}, where the black crosses show the median accuracies in the correctly-estimated case for comparison. We see that the accuracies (orange circles) are hardly affected by the underestimation. The precisions (black diamonds), in contrast, are now smaller than the accuracies for both the position and the velocity, by a factor of 2.3 on average for the positions (range of 2.0 to 2.6), and a factor of
2.6 for the velocities (range 2.3 to 2.9). Hence, underestimating the measurement uncertainties affects the predicted uncertainties (precisions) in the inferred parameters, but not their accuracies. Putting it another way: not knowing the measurement accuracy exactly will not affect how well we can actually determine the spacecraft position and velocity.

\subsection{Other variations}\label{sec:other_variations}

% All done for S/C moving at relativistic speeds, 0-0.5c, true SSB time 10-20 yr.

The results presented in the previous section were for the spacecraft 0.1--10\,ly from the SSB (average over 100 randomly selected distances in this range; see section~\ref{sec:truepars}).  Further tests indicate that the navigation performance is independent of this spacecraft distance. Specifically, if we reduce this distance range by a factor of 300, with everything else kept fixed, the performance is unchanged. Of course, if the spacecraft moved so far that all stars appeared in the same small region of the sky, then the performance would be affected.

In contrast, performance does degrade if we use more-distant stars. For example, if we use 20 stars spread uniformly in space out to 1\,\kpc\ (3260\,\ly), then with angular distance measurements of 1\as\ accuracy the spacecraft positional accuracy is reduced by a factor of a few hundred.
% SSB time accuracy also reduced by same order of magnitude.
We might attribute this to the fact that the stars now have smaller parallaxes (the spacecraft still moves only 0.1--10\,ly from the SSB). Yet the typical parallax is of order 600\as, still much larger than our 1\as\ measurement accuracy. Moreover, the previous test of reducing instead the distance of the spacecraft reduced the stars' parallaxes by the same order-of-magnitude, yet this did not affect performance. Hence the signal-to-noise ratio of the measured angles (the term in square brackets in equation~\ref{eqn:likelihood_angdist}) cannot be the reason for the performance degradation.
%%% This is quite counter-intuitive, but results have been double-checked. More analysis required.
Note that the spacecraft velocity accuracy degrades by a factor of just 1.5, which would be consistent with the distance-independence of aberration.
% Yet putting all 20 stars at 1\,\kpc\ degrades the velcocity accuracy by a couple of orders of magnitude, and the position accuracy by a couple of orders of magnitude more. Interestingly, adding radial velocities with an accuray of 10\,\kms\ recovers a lot of this performance.

Additional tests suggest there is a complicated dependence of the performance on both the average distance to the stars and their distance distribution (and also whether we include radial velocity measurements). Exploring this is of interest to better understand how the method works, although in practice we can just use the nearest stars, as this appears to give the best performance.

%%%%%%%%%%%%%%%%%%%%%%%%%%%%%%%%%%%%%%%%%%%%%%%%%%%%%%%%%%%%%%%%%%%%%%%
\section{Summary, discussion, and conclusions}\label{sec:conclusions}

I have developed and demonstrated a scheme to navigate a spacecraft in interstellar space using a catalogue of the 3D positions and 3D velocities of nearby stars. The scheme exploits the parallax and aberration of the stars, both of which depend on the position and velocity of the spacecraft, by making measurements only of the angular distances between stars. Using multiple stars we can untangle the aberrational and parallactic contributions to the observed angular shifts to infer the 3D position and 3D velocity of the spacecraft relative to the SSB.
With the 20 nearest stars and an onboard measurement accuracy of 1\as, I have shown via simulations that the position and velocity of the spacecraft can be determined to within 3\,\au\ and 2\,\kms\ respectively.
%%% N=10 with just angdist is poorer than the 1/sqrt(N-1) trend, due to near degeneracy.
%With just 10 stars, i.e.\ 9 angular measurements, and an onboard measurement accuracy of 1\as, I have shown via simulations that the position and velocity of the spacecraft can be determined to within 11\,\au\ and 9\,\kms\ respectively.
Increasing this to 100 stars improves the accuracies to 1.3\,\au\ and 0.7\,\kms\ respectively. The accuracy improves approximately as the inverse of the square root of the number of stars.  The navigational accuracies are found to be in direct proportion to the measurement accuracy: With 20 stars and measurement accuracies of 1\,\as, 0.1\as, and 0.001\as, we achieve positional accuracies of 3\,\au, 0.3\,\au, and 0.003\,\au\ respectively, and velocity accuracies of 2\,\kms, 0.2\,\kms, and 0.002\,\kms\ respectively.  As aberration is a large effect -- more than 1\as\ for velocities above 1.5\,\kms\ -- the accuracy of the velocity determination is essentially independent the spacecraft velocity, so is as good for relativistic as for non-relativistic spacecraft.

The method uses MCMC to sample the likelihood (formally the posterior with a uniform prior), and from the resulting set of samples we can estimate the uncertainties in the inferred parameters. We find these to be close to the amplitudes of the residuals in general, making them a useful measure of the accuracy of the inferred parameters in a real-world situation.

We may also measure stellar radial velocities from the spacecraft, as these too encode information about both the position and velocity of a star when compared to the catalogue.  Whether these are useful depends on their accuracies.  Using only radial velocities with an accuracy of 10\,\kms\ gives poor results: with 20 stars the positional accuracy is 160\,\au\ and the velocity accuracy is 10\,\kms. Improving the measurement accuracy by some factor improves the accuracy of the inferred positions and velocities by the same factor, as was the case with angular distance measurements.

Combining the two types of measurements may improve the accuracy attainable from either alone. For example, with 20 stars and 1\as\ angular distances, adding radial velocities of 1\,\kms\ accuracy improves the position and velocity accuracies on average by factors of 2.8 and 4.5 respectively.
% From Ntruepar=20
% Units are: position [au] speed [km/s]
% Accuracy both:      1.22e+00  5.49e-01 
% Accuracy angdist:  3.46e+00  2.53e+00
But if the radial velocities are only accurate to 10\,\kms, the positional accuracy only improves by 10\% and the velocity accuracy not at all.
% From Ntruepar=100
% Units are: position [au] speed [km/s]
% Accuracy both:     2.80  2.28
% Accuracy angdist: 3.06 2.26 
As it may be difficult to achieve stellar radial velocities more accurate than a few \kms\ from an interstellar spacecraft, but it is comparatively easy to measure angles to 1\as, attaining higher navigation accuracies should focus on achieving more accurate angular measurements.  Additional stars could be used, but the performance only improves as $1/\sqrt{N}$, in line with expectations.

Due to the motion of the stars, and the assumption that a relativistic spacecraft could not keep track of SSB time, we were forced to formally include the measurement time as a seventh unknown parameter in the inference. This has only a small impact on the inference, however, and consequently cannot be inferred very accurately. SSB time could be estimated by sending a time signal from the Earth, which requires the spacecraft to correct for the signal travel time based on its current position. But for the same reason that time cannot be inferred accurately, this improved knowledge is not expected to improve the navigation performance by much.
  Such signals could also be used to provide updates of the star catalogue, although with a proper motion accuracy of 0.02\,\mas/\yr\ already being attained by Gaia, the catalogue positions would only degrade by 2\,\mas\ in a century.

The main results were computed for spacecraft placed randomly in space up to 10\,ly from the Sun with velocities up to 0.5c.  The first interstellar missions are likely to be local, so we can use just nearby stars for this navigation.  For more distant sojourns, the best performance would probably be obtained by using those stars expected to be nearest to the spacecraft, and surrounding it reasonably isotropically in its rest frame.  The navigation should continue to work just as well for spacecraft at least as far as the average distance of accurately-measured stars in the star catalogue which, with Gaia, is hundreds of light years. The position and velocity accuracies are likewise independent of the velocity of the spacecraft (relative to the SSB), although at extreme relativistic velocities some degradation will eventually occur due to strong aberration making most stars lie in a small part of the sky as seen from the spacecraft.

%The main point of this article was a proof of concept of parallactic/aberrational navigation, to demonstrate that it works and to determine what navigational accuracy can be achieved with what inputs.  Higher accuracy may be achieved with pulsar timing observations. The method presented here can still play a role, perhaps as an initialization for pulsar navigation, or as a comparison or backup.

This study is primarily conceptual. Although we only rely on measurements that could be made from a relativistic spacecraft, e.g.\ we have not assumed the availability of a fixed reference frame, we have not considered the instruments themselves.  Angular distances could be measured using a highly accurate sextant, which is similar in principle to the astrometric instruments on Hipparcos (1\,\mas\ accuracy) and Gaia (a few \muas\ accuracy), except that these have a fixed ``basic angle'' between the two fields-of-view; they then allow stars to drift over the observing field, essentially converting time differences between focal plane crossings into angular distances, similar to how ground-based meridian circles operate. As 1\as\ can easily be achieved by direct-imaging commercial star trackers, it seems reasonable to assume that a specially-designed space sextant could do a lot better.  Random errors can easily be beaten down through multiple measurements: observation time (or photons) is hardly an issue for a decades-long mission. The limiting factor will be systematic errors.  Whether we can get to Hipparcos or Gaia accuracies depends strongly on the size of the spacecraft, and so on what metrology can be introduced to determine the basic angle.
%A particular issue here is ensuring the long-term reliability of the instrument.

Some other implementational issues have not been considered. An important one is binary stars. I have assumed that the star catalogue gives us whatever information we need to extrapolate the positions and velocities of stars over decades. The accuracy of Gaia parallaxes and proper motion, combined with ground-based radial velocities, is sufficiently high to ensure this is possible for single stars. But stars in compact binary systems can have large enough accelerations that the assumption of linear motion is inadequate. Either we need to model their accelerations, or we must exclude them from the catalogue.

%The second issue is that the observed Doppler shift is generated by more than just the radial motion of the star.  The main additional factors are the gravitational redshift (of order 0.6\,\kms\ for the Sun) and the convective blueshift (of order 0.1\,\kms).  When using radial velocities at this order of accuracy, these effects would need to be accommodated.

\section*{Acknowledgements}

I would like to thank Alex Bombrun (ESAC) for comments on a draft manuscript.

\section*{References}

Bailer-Jones C.A.L.\ 2015,
%{\em Estimating distances from parallaxes},  
PASP, 127, 994

Becker W., Bernhardt M.G., Jessner A.\ 2013,
%{\em Autonomous spacecraft navigation with pulsars},
Acta Futura, 7, 1
% 1--28
% https://arxiv.org/abs/1305.4842v1

Butkevich A.G., Klioner S.A.\ 2008,
%{\em Determination of the barycentric velocity of an astrometric satellite using its own observational data},
in Proceedings of the International Astronomical Union
Symposium S248: A Giant Step: from Milli- to Micro-arcsecond Astrometry (Cambridge University Press),
252
% --255
% I do not know who the editors are. It is not even on the CUP website.
% https://ui.adsabs.harvard.edu/abs/2008IAUS..248..252B/abstract
% https://doi.org/10.1017/S1743921308019194

Butkevitch A.G., Lindegren L.\ 2014,
%{\em Rigorous treatment of barycentric stellar motion},
A\&A, 570, A62

Calabro' E.\ 2011,
%{\em Relativistic aberrational interstellar navigation},
%Acta Astronautica,
AcAau
69, 360
%--364

Christian J.\ 2019,
%{\em StarNAV: Autonomous optical navigation of a spacecraft by the relativistic perturbation of starlight},
Sensors, 19, 4064

Foreman-Mackey D., Hogg D.W., Lang D., Goodman J.\ 2013,
% {\em emcee: The MCMC hammer},
PASP, 125, 306 

Gaia Collaboration 2016,
%{\em The Gaia mission},
A\&A, 595, A1

Gaia Collaboration 2018,
% {\em Gaia Data Release 2. Summary of the contents and survey properties},
A\&A, 616, A1

Goodman J., Weare J.\ 2010,
%{\em Ensemble samplers with affine invariance},
Comm.\ App.\ Math.\ and Comp.\ Sci., 5, 65

Hoag D.G., Wrigley W.\ 1975,
% {\em Navigation and guidance in interstellar space},
%Acta Astronautica,
AcAau,
2, 513
%--533

Iess L., Di Benedetto M., James N., Mercolino M., Simone L., Tortora P.\ 2014,
%{\em Astra: Interdisciplinary study on enhancement of the end-to-end accuracy for spacecraft tracking techniques},
%Acta Astronautica,
AcAau,
94, 699
%--707

James N., Abellob R., Lanucarab M., Mercolinob M., Madd\'e R.\ 2009,
%{\em Implementation of an ESA delta-DOR capability},
%Acta Astronautica,
AcAau,
64, 1041--1049

Klioner S.A.\ 2003,
% {\em A practical relativistic model for microarcsecond astrometry in space},
AJ, 125, 1580

Moskowitz S., Devereu W.P.\ 1968,
% {\em Trans-stellar space navigation},
% American Institute of Aeronautics and Astronautics Journal
AIAAJ,
6, 1021--1029

Perryman M.A.C., ed.\ 1997, {\em The HIPPARCOS and Tycho catalogues}, (Paris: ESA SP), 1200

Shemar S., et al., 2016,
% {\em Towards practical autonomous deep-space navigation using X-Ray pulsar timing},
%Experimental Astronomy,
ExA,
42, 101
%-–138

Wenger M., Ochsenbein F., Egret D., et al.\ 2000,
%{\em The SIMBAD astronomical database. The CDS reference database for astronomical objects},
% Astronomy \& Astrophysics Supplement,
AAS
143, 9
%--22

\end{document}